\documentclass[twocolumn]{autart}

\usepackage{cite}

\usepackage{amsmath,amssymb,amsfonts}
\usepackage{textcomp}
\usepackage[dvipsnames]{xcolor}
\usepackage{graphicx}
\usepackage{subfig}
\usepackage{epstopdf}
\usepackage{comment}
\usepackage{caption}
\usepackage{mathptmx}

\theoremstyle{description}
\newtheorem{lemm}{Lemma}
\newtheorem{thme}{Theorem}

\newtheorem{prop}{Proposition}
\newtheorem{ass}{Assumption}

\newtheorem{rmk}{Remark}

\allowdisplaybreaks

\definecolor{darkgreen}{rgb}{0.0,0.6, 0.0}

\begin{document}

\begin{frontmatter}

\title{Performance-Barrier Event-Triggered Control of a 
Class of Reaction-Diffusion PDEs\thanksref{t1}}

\thanks[t1]{A preliminary version of this work appears in \cite{rathnayake2023pfmnc}. The work of B. Rathnayake and  M. Diagne was funded by the NSF CAREER Award CMMI-2302030 and  the NSF grant CMMI-2222250. The work of J. Cort\'es was funded by AFOSR Award FA9550-23-1-0740. The work of M. Krstic was funded by the NSF grant ECCS-2151525 and the AFOSR grant FA9550-23-1-0535.}

\thanks[footnoteinfo]{Corresponding author B.~Rathnayake Tel. +1(518)5965193.}

\author[Baiae]{Bhathiya Rathnayake\thanksref{footnoteinfo}}\ead{brm222@ucsd.edu},    
\author[troy]{Mamadou Diagne}\ead{mdiagne@ucsd.edu },               
\author[troy]{Jorge Cortes}\ead{cortes@ucsd.edu},  
\author[troy]{Miroslav Krstic}\ead{krstic@ucsd.edu}

\address[Baiae]{Department of Electrical and Computer Engineering, University of California, San Diego, La Jolla, CA 92093, USA}  
\address[troy]{Department of Mechanical and Aerospace Engineering, University of California, San Diego, La Jolla, CA 92093, USA}

\begin{abstract}
We employ the recent performance-barrier event-triggered control (P-ETC) for achieving global exponential convergence of a class of reaction-diffusion PDEs via PDE backstepping control.  Rather than insisting on a strictly monotonic decrease of the Lyapunov function for the closed-loop system, P-ETC allows the Lyapunov function to increase as long as it remains below an acceptable performance-barrier. This approach integrates a performance residual—the difference between the value of the performance-barrier and the Lyapunov function—into the triggering mechanism. The integration adds flexibility and results in fewer control updates than with \textit{regular} ETC (R-ETC) that demands a monotonic decrease of the Lyapunov function. Our P-ETC PDE backstepping design ensures global exponential convergence of the closed-loop system solution to zero in the spatial   $L^2$ norm, without encountering Zeno phenomenon. To avoid continuous monitoring of the triggering function that generates events, we develop periodic event-triggered and self-triggered variants (P-PETC and P-STC, respectively) of the P-ETC. The P-PETC only requires periodic evaluation of the triggering function whereas the P-STC preemptively computes the time of the next event at the current event time using the system model and continuously available system states. The P-PETC and P-STC also ensure a Zeno-free behavior and deliver performance equivalent to that of the continuous-time P-ETC which requires continuous evaluation of the triggering function, in addition to the continuous sensing of the state. We provide numerical simulations to illustrate the proposed technique and to compare it with R-ETC associated with strictly decreasing Lyapunov functions.
\end{abstract}

\end{frontmatter}

\section{Introduction}
\subsection{The state of the Art}
Event-triggered control (ETC) introduces an alternative to typical sampled-data control that updates the control input based on a fixed periodic or aperiodic sampling schedule. Distinctively, ETC updates the control input only after the occurrence of certain events generated by an appropriate triggering mechanism that depends on system states \cite{heemels2012introduction}. As such, ETC can be construed as a form of sampled-data control that seamlessly integrates feedback into control update processes. By leveraging the power of feedback, ETC allows control input to be updated only when necessary, reducing control updates, while still maintaining a satisfactory closed-loop system performance.

Generally, ETC is built on two primary elements: a feedback control law, which ensures desired closed-loop system properties, and an event-triggering mechanism, which determines when control input should be updated. To function effectively, ETC must stave off \textit{Zeno behavior}—the scenario of occurrence of an infinite number of control updates within a finite time span. This is usually achieved via careful design of the event triggering mechanism, guaranteeing a \textit{minimum dwell-time} \textit{i.e.,} a uniform positive lower bound between two consecutive events. In the recent years, an array of impressive results related to ETC have been presented for systems governed by both linear and nonlinear ODEs \cite{heemels2012introduction,mazo2008event,girard2014dynamic,tallapragada2015event,liu2015small,taylor2020safety}. This has spurred exploration into ETC strategies for systems described by PDEs \cite{espitia2016event,espitia2020observer,espitia2019event,katz2020boundary,rathnayake2021observer,rathnayake2022sampled,wang2022eventa,rathnayake2022observer,lhachemi2024event,ji2023hyper}. Of particular relevance to our current study are \cite{rathnayake2021observer} and \cite{rathnayake2022sampled}, which propose event-triggered boundary control strategies for a class of reaction-diffusion PDEs, using dynamic event-triggers under anti-collocated and collocated boundary sensing and actuation.

One significant limitation of ETC strategies, both for ODE and PDE systems, lies in the necessity for continuous monitoring of event-triggering functions to detect events. This does not lend itself well to digital implementation. For these strategies, we use the term \textit{continuous-time event-triggered control (CETC)}. To address this issue, one approach involves designing a periodically-checked event-triggering condition, leading to what is commonly referred to as \textit{periodic event-triggered control (PETC)} \cite{heemels2012periodic}. Here, the event-triggering condition is checked at regular intervals, and at each instant, a decision is made regarding the necessity of a new control update. Another solution is \textit{self-triggered control (STC)} \cite{heemels2012introduction}, which preemptively calculates next event time at current event time, based on predictions utilizing state measurements and insights into plant dynamics. Both PETC and STC maintain the resource efficiency of CETC. With PETC, the control input is updated aperiodically at events, even though the triggering condition is checked periodically. Similarly, in STC, the control input is updated aperiodically at events, while also computing the time of the next event. This makes both types of control amenable to digital implementations, as the triggering condition or next event time can be processed in standard time-sliced digital software. During the past few years, several interesting works devoted to both PETC \cite{heemels2012periodic,heemels2013model,borgers2018periodic,fu2018decentralized,wang2019periodic,linsenmayer2019periodic,seidel2023window} and STC \cite{mazo2008event,mazo2009self,anta2010sample,yi2018dynamic,yang2019self,wan2020dynamic,cao2023self} of ODE systems have been published. To the best of our knowledge, only the studies \cite{selivanov2016distributed,wakaiki2020event,wakaiki2019stability,rathnayake2023observertac,rathnayake2023observerstefan} have explored STC or/and PETC strategies for infinite-dimensional systems. In \cite{selivanov2016distributed}, the authors develop a PETC approach for a network of semilinear diffusion PDEs with \textit{in-domain} actuation and distributed or point measurements. Leveraging semigroup theory, \cite{wakaiki2020event} discusses a full-state feedback PETC mechanism for infinite-dimensional systems faced with unbounded control operators, while \cite{wakaiki2019stability} introduces a full-state feedback STC approach for infinite-dimensional systems with bounded control operators. In \cite{rathnayake2023observertac}, the authors present the first PETC and STC PDE backstepping \textit{boundary control} design for a class of reaction-diffusion PDEs using \textit{boundary measurements}. The observer-based PETC extension to diffusion PDEs with moving boundaries featuring the one-phase Stefan problem is discussed in \cite{rathnayake2023observerstefan}. 

\subsection{Contributions} 

In this study, we employ the recently introduced \textit{performance-barrier event-triggered control (P-ETC)} \cite{ong2021performance} for control of a class of reaction-diffusion PDEs via PDE backstepping control. This novel approach offers significantly longer dwell-times between events compared to the recently devised dynamic event-triggered boundary control method for a class of reaction-diffusion PDEs \cite{rathnayake2021observer,rathnayake2022sampled}. The contributions of this paper are threefold. Before delving into their specifics, it is essential to clarify the terminology, particularly the term \textit{performance-barrier} taken from~\cite{ong2021performance}, which forms the crux of our novel approach.

The term \textit{performance-barrier} is inspired by the safety-critical control literature \cite{ames2016control,ames2019control,taylor2020safety,krstic2023inverse}, although our context does not directly deal with safety. In our study, \textit{performance} refers to the nominal decrease of the Lyapunov function, which serves as a measure of system convergence. On the other hand, \textit{barrier} alludes to a boundary or threshold that the system should ideally not cross. Together, the \textit{performance-barrier} terminology encapsulates the idea of comparing the Lyapunov function's behavior to an acceptable nominal decrease, treating it as a boundary that should not be violated. 

 In \cite{rathnayake2021observer,rathnayake2022sampled}, the triggering mechanism forces a monotonic decrease in the Lyapunov function of the closed-loop system by ensuring its time derivative remains strictly negative. In terms of performance, this guarantees the Lyapunov function decreases faster than a specific exponentially decaying signal, which incorporates the initial data. We refer to this latter signal as the \textit{performance-barrier}. Our hypothesis is that by allowing some flexibility in the Lyapunov function's behavior—letting it deviate from a strict monotonic decrease but still respecting the performance-barrier—we can achieve longer dwell-times between events leading to less frequent control updates. This idea was carried out in the work of Ong \textit{et al.} \cite{ong2021performance} for ODEs, which develops an event-triggered control design that integrates both the time derivative and the value of the Lyapunov function into the triggering criterion.

To operationalize this idea, we utilize the concept of a \textit{performance residual} \cite{ong2021performance}, which represents the difference between the value of the performance-barrier and the Lyapunov function. By incorporating this residual into the triggering mechanism, we allow the Lyapunov function greater flexibility in its behavior, eliminating the need for it to decrease monotonically at all times. Given that the triggering function is continuously monitored, this methodology is termed performance-barrier continuous-time event-triggered control (P-CETC). Meanwhile, we label the strategies introduced in \cite{rathnayake2021observer,rathnayake2022sampled} as \text{regular} continuous-time event-triggered control (R-CETC) where the term \textit{regular} is used to refer to the strictly decreasing nature of the associated Lyapunov functions. Notably, the P-CETC offers longer dwell-times at any given state compared to the R-CETC. Importantly, this is achieved while excluding Zeno behavior from the closed-loop system and still maintaining adherence to the performance-barrier which leads to the global exponential convergence of the closed-loop system solution to zero in the spatial   $L^2$ norm. The design of the P-CETC and analysis of its closed-loop system properties are our first contribution.

Moreover, we develop periodic and self-triggered variants of the P-CETC that ensure performance equivalence with the original P-CETC. These are inspired by the regular PETC (R-PETC) and regular STC (R-STC) proposed in \cite{rathnayake2023observertac} to overcome the need of continuous monitoring of the triggering functions of the R-CETC \cite{rathnayake2021observer,rathnayake2022sampled}. The performance-barrier periodic event-triggered control (P-PETC) is the result of a careful redesign of the continuous-time event-triggering function, modified to necessitate only periodic evaluations. We derive a periodic event triggering function by determining an upper bound on the underlying continuous-time event-triggering function between two consecutive periodic evaluations. This leads to the specification of the explicit maximum allowable sampling diameter for P-PETC strategy. Since the event-triggering function is evaluated periodically, it inherently prevents Zeno behavior. The proposed performance-barrier self-triggered control (P-STC) strategy involves designing a state dependent function with a uniform and positive lower bound. Evaluating this function at the current event time provides the time duration until the next event, based on the system states. This function is constructed using upper and lower bounds on the variables of the original continuous-time event trigger. Since the function maintains a uniform positive lower bound, the closed-loop system is inherently Zeno-free. Our second and third contributions are, respectively, the design and analysis of the closed-loop system properties of the P-PETC and P-STC. 

\begin{figure}
    \centering
    \includegraphics[width=1\linewidth]{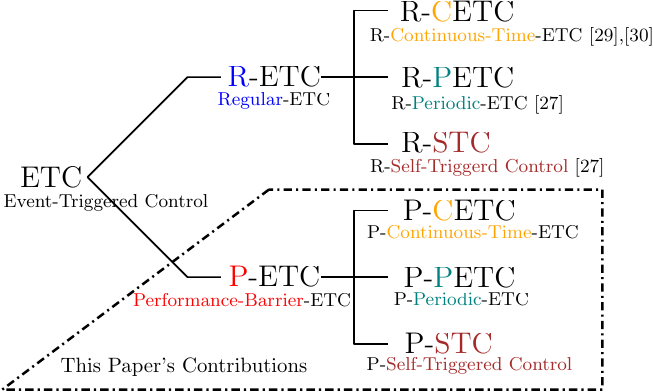}
    \caption{Overview of the designs discussed in this paper.}
    \label{abbrev}
\end{figure}

In fact, the proposed P-CETC, P-PETC, and P-STC generalize the full-state feedback versions of the R-CETC, R-PETC, and R-STC introduced in \cite{rathnayake2021observer,rathnayake2022sampled,rathnayake2023observertac} by incorporating Lyapunov functions that are not strictly decreasing. Furthermore, we have developed designs for handling a class of reaction-diffusion PDEs with a spatially varying reaction coefficient, which was not addressed in \cite{rathnayake2021observer,rathnayake2022sampled,rathnayake2023observertac}. Without major modifications, the P-ETC triggering mechanisms proposed in this paper are also applicable to full-state feedback control of coupled hyperbolic PDEs \cite{espitia2020observer}, diffusion PDEs with moving boundaries \cite{rathnayake2022observer}, and PDE-ODE cascades \cite{wang2022eventa,ji2023hyper}. Fig. \ref{abbrev} provides an overview of the designs. A preliminary version of the P-CETC design for a constant parameter reaction-diffusion PDE appears in our conference paper \cite{rathnayake2023pfmnc}.

\subsection{Organization}
The rest of the paper is organized as follows. In Section 2, we present the continuous-time control and its emulation. Section 3 presents the P-ETC designs that comprise the P-CETC, P-PETC, and the P-STC. A numerical example is provided in Section 4 to illustrate the results, and conclusions are provided in Section 5.        

\subsection{Notation}
$\mathbb{R}_+$ is the positive real line whereas $\mathbb{N}$ is the set of natural numbers including zero.  By $C^{0}(A;\Omega)$, we denote the class of continuous functions on $A\subseteq\mathbb{R}^{n}$, which takes values in $\Omega\subseteq\mathbb{R}$. By $C^{k}(A;\Omega)$, where $k\geq 1$, we denote the class of continuous functions on $A$, which takes values in $\Omega$ and has continuous derivatives of order $k$.  $L^{2}(0,1)$ denotes the equivalence class of Lebesgue measurable functions $f:[0,1]\rightarrow\mathbb{R}$ such that $\Vert f\Vert=\big(\int_{0}^{1}\vert f(x)\vert^{2}\big)^{1/2}<\infty$. Let $u:[0,1]\times\mathbb{R}_{+}\rightarrow\mathbb{R}$ be given. $u[t]$ denotes the profile of $u$ at certain $t\geq 0$, \textit{i.e.,} $\big(u[t]\big)(x)=u(x,t),$ for all $x\in [0,1]$. For an interval $J\subseteq\mathbb{R}_{+},$ the space $C^{0}\big(J;L^{2}(0,1)\big)$ is the space of continuous mappings $J\ni t\rightarrow u[t]\in L^{2}(0,1)$. 

\section{Continuous-time Control and Emulation}
Let us consider the following one-dimensional reaction-diffusion PDE with a spatially varying reaction coefficient
\begin{align}\label{ctpe1}
u_{t}(x,t)&=\varepsilon u_{xx}(x,t)+\lambda(x) u(x,t),\text{ }\forall x\in (0,1),\\
\label{ctpe2}
\theta_1u_{x}(0,t)&=-\theta_2u(0,t),\\\label{ctpe3}
u_{x}(1,t)&=-qu(1,t)+U(t),
\end{align}
for all $t>0$. The plant parameters $\varepsilon$, $\lambda(x)$, and $q$ are such that $\varepsilon, q > 0$, and $\lambda \in C^2([0,1];\mathbb{R}_+)$, and the coefficients $\theta_1, \theta_2$ satisfy $\theta_1\theta_2=0,\theta_1+\theta_2=1$. The system state is defined as $u: [0,1] \times [0,\infty) \rightarrow \mathbb{R}$, and $U(t)$ is the continuous-time control input. The initial condition is such that $u[0] \in L^{2}(0,1)$. Note that $\theta_1$ and $\theta_2$ are either $0$ or $1$, and $\theta_1 \neq \theta_2$. The case $\theta_1 = 1$, $\theta_2 = 0$ leads to a Neumann boundary condition at $x = 0$, whereas the case $\theta_1 = 0$, $\theta_2 = 1$ leads to a Dirichlet boundary condition at $x = 0$. 

\begin{rmk}\rm 
Consider the following one-dimensional linear reaction-advection-diffusion PDE with spatially varying diffusion $\bar{\varepsilon}(\bar{x})$, advection $\bar{b}(\bar{x})$, and reaction $\bar{\lambda}(\bar{x})$ coefficients:
\begin{align}
\begin{split}
\bar{u}_t(\bar{x},t) &= \bar{\varepsilon}(\bar{x}) \bar{u}_{\bar{x}\bar{x}}(x,t) + \bar{b}(\bar{x}) \bar{u}_{\bar{x}}(\bar{x},t) + \bar{\lambda}(\bar{x}) \bar{u}(\bar{x},t),\\&\qquad\qquad\qquad\qquad\qquad\quad\qquad\text{ }\forall \bar{x}\in(0,1), \label{nnn1}\end{split} \\
\bar{u}(0,t) &=0,  \\
\bar{u}_{\bar{x}}(1,t)&= -\bar{q}\bar{u}(1,t)+\bar{U}(t),\label{nnn2}
\end{align}
for all $t > 0$ with an initial condition such that $\bar{u}[0] \in L^{2}(0,1)$ and a boundary control input $\bar{U}(t)$, where $\bar{\varepsilon}\in C^4([0,1];\mathbb{R}_+), \bar{\lambda} \in C^2([0,1];\mathbb{R}_+)$, $\bar{b} \in C^3([0,1];\mathbb{R})$, and $\bar{q} > 0$. Using a so-called gauge transformation (see Section 4.8 of \cite{krstic2008boundary}), the system \eqref{nnn1}-\eqref{nnn2} can be transformed to \eqref{ctpe1}-\eqref{ctpe3}, with $\theta_1=0$ and $\theta_2=1$. Therefore, the results developed in this paper are applicable to a more general linear parabolic PDE having the form \eqref{nnn1}-\eqref{nnn2}.
\end{rmk}

In this section, we briefly present the continuous-time PDE backstepping boundary control design for the system \eqref{ctpe1}-\eqref{ctpe3}. This is followed by its emulation for event-triggered boundary control.

\subsection{Backstepping Control Design}

\begin{ass}\label{ass1} The parameters $q,\varepsilon>0,\lambda\in C^2([0,1];\mathbb{R}_+)$ satisfy the following relation:\begin{equation}
q>\frac{\lambda_{max}}{2\varepsilon}+\frac{\theta_1}{2},
\end{equation}
where 
\begin{equation}
\lambda_{max}\triangleq\max_{x\in [0,1]}\lambda(x).
\end{equation}
\end{ass}

 Assumption \ref{ass1} is important in ensuring the stability of the target system under PDE backstepping control with 
 dynamic event-triggering. This is because we intentionally avoid using the signal $u(1, t)$ in the nominal control law. Such avoidance is crucial for dynamic ETC design due to the challenges associated with obtaining a meaningful bound on the rate of change of $u(1, t)$. However, an eigenfunction expansion of the solution of \eqref{ctpe1}-\eqref{ctpe3} with $U(t)=0$ shows that the system is unstable when $\min_{x\in [0,1]}\lambda(x)>\varepsilon\pi^2/4^{\theta_1}$, no matter what $q>0$.

Consider the invertible backstepping transformation
\begin{equation}\label{ctbtd}
w(x,t)=u(x,t)-\int_{0}^{x}K(x,y)u(y,t)dy,
\end{equation}
defined in the domain $0\leq y\leq x\leq 1$, where $K(x,y)$ satisfies
\begin{align}\label{ctcke1}
K_{xx}(x,y)- K_{yy}(x,y)&=\frac{\lambda(y)}{\varepsilon}K(x,y),
\\
\theta_1K_{y}(x,0)&=-\theta_2K(x,0),
\\
K(x,x)&=-\frac{1}{2\varepsilon}\int_{0}^x\lambda(y)dy,\label{aamlper}
\end{align}
 for $0\leq y\leq x\leq 1$. Then, using standard arguments in PDE backstepping boundary control, we can show that  the transformation \eqref{ctbtd}-\eqref{aamlper} and a control law $U(t)$ chosen as\begin{equation}\label{ctcl}
U(t)=\int_{0}^{1}k(y)u(y,t)dy,
\end{equation}
where 
\begin{equation}\label{ase1}
    k(y)=\wp K(1,y)+K_{x}(1,y),
\end{equation}
with
\begin{equation}\label{rt}
\wp=q-\frac{1}{2\varepsilon}\int_{0}^{1}\lambda(y)dy,
\end{equation}
map the system \eqref{ctpe1}-\eqref{ctpe3} into the following target system:
\begin{align}\label{etotse1}
w_{t}(x,t)&=\varepsilon w_{xx}(x,t),
\\\label{etotse2}
\theta_1w_x(0,t)&=-\theta_2w(0,t),
\\\label{etotse3}
w_{x}(1,t)&=-\wp w(1,t).
\end{align}
The inverse transformation of \eqref{ctbtd}-\eqref{aamlper} is given by
\begin{equation}\label{puoi}
u(x,t)=w(x,t)+\int_{0}^{x}L(x,y)w(y,t)dy,
\end{equation}
where $L(x,y)$ satisfies
\begin{align}\label{ctckeg}
L_{xx}(x,y)- L_{yy}(x,y)&=-\frac{\lambda(y)}{\varepsilon}L(x,y),
\\
\theta_1L_{y}(x,0)&=-\theta_2L(x,0),
\\
L(x,x)&=-\frac{1}{2\varepsilon}\int_{0}^x\lambda(y)dy,\label{kkllspip}
\end{align}
for $0\leq y\leq x\leq 1$. Following arguments similar to those provided in \cite{liu2003boundary},\cite{smyshlyaev2004closed}, it can be shown that the PDE systems \eqref{ctcke1}-\eqref{aamlper} and \eqref{ctckeg}-\eqref{kkllspip} admit unique $C^3$ solutions in the domain $0\leq y\leq x\leq 1$, provided that $\lambda\in C^2([0,1])$.

\subsection{Emulation of the Backstepping Boundary Control}
Our objective is to drive the plant \eqref{ctpe1}-\eqref{ctpe3} to zero in the spatial   $L^2$ norm by sampling the continuous-time controller $U(t)$ given by \eqref{ctcl} at a certain sequence of time instants $\{t_{j}\}_{j\in\mathbb{N}}$. The characterization of these time instants will be based on several dynamic event triggers, which we will detail later. The control input is held constant between two successive time instants and is updated when a certain condition is met. We define the control input for $t\in[t_{j},t_{j+1}),j\in\mathbb{N}$ as
\begin{equation}\label{etcla}
U_{j} \triangleq U(t_{j})=\int_{0}^{1}k(y)u(y,t_{j})dy.
\end{equation}Accordingly, the boundary condition \eqref{ctpe3} is modified as follows:\begin{equation}\label{mctpe3x}
u_{x}(1,t)+qu(1,t)=U_{j},
\end{equation}
for $t\in[t_{j},t_{j+1}),j\in\mathbb{N}$. The deviation between the continuous-time control law and its sampled counterpart, referred to as the input holding error, is defined as
\begin{equation}\label{dt}
d(t) \triangleq\int_{0}^1 k(y)\big(u(y,t_j)-u(y,t)\big)dy,
\end{equation} for $t\in[t_{j},t_{j+1}),j\in\mathbb{N}$. It can be shown that the backstepping transformation \eqref{ctbtd}-\eqref{aamlper} applied on the system \eqref{ctpe1},\eqref{ctpe2},\eqref{etcla},\eqref{mctpe3x} between $t_{j}$ and $t_{j+1}$, yields the following target system, valid for $t\in[t_{j},t_{j+1}),j\in\mathbb{N}$:

\begin{align}\label{tup}
w_{t}(x,t)&=\varepsilon w_{xx}(x,t),
\\
\theta_1w_x(0,t)&=-\theta_2w(0,t),
\\\label{tup3}
w_{x}(1,t)&=-\wp w(1,t)+d(t).
\end{align}

Below we present the well-posedness of the closed-loop system \eqref{ctpe1},\eqref{ctpe2},\eqref{etcla},\eqref{mctpe3x} between two consecutive sampling instants $t_{j}$ and $t_{j+1}$.


\begin{prop}\textnormal{\textbf{({Well-Posedness between control updates)}}}\label{cor1}
For every $u[t_{j}]\in L^{2}(0,1)$, there exist a unique solution $u:[t_j,t_{j+1}]\times[0,1]\rightarrow\mathbb{R}$ between two time instants $t_{j}$ and $t_{j+1}$ such that $u\in C^0\big([t_j,t_{j+1}];L^2(0,1)\big)\cap C^1\big((t_j,t_{j+1})\times [0,1]\big)$ with $u[t]\in C^2([0,1])$ which satisfy \eqref{ctpe2},\eqref{mctpe3x} for $t\in(t_j,t_{j+1}]$ and \eqref{ctpe1} for $t\in(t_j,t_{j+1}],x\in(0,1)$.  
\end{prop}

Proposition \ref{cor1} is a straightforward application of Theorem 4.11 in \cite{karafyllis2019input}.  This will be used through out the paper to establish the well-posedness of the closed-loop system in the hybrid time domain $\mathbb{T}=\bigcup_{j=0}^{J-1}[t_j,t_{j+1}]\times\{j\}$ where $\mathbb{T}\subset \mathbb{R}_{\geq 0}\times \mathbb{N}$ and $J$ is possibly $\infty$ and/or $t_J=\infty$. 

\section{Performance-barrier Event-triggered Control}
In this section, we introduce the design of the performance-barrier event-triggered control (P-ETC) across three configurations: continuous-time event-triggered, periodic event-triggered, and self-triggered control (P-CETC, P-PETC, and P-STC, respectively). Before delving into P-ETC, we will briefly discuss the regular continuous-time event-triggered control (R-CETC), which is analogous to the designs introduced in \cite{rathnayake2021observer, rathnayake2022sampled}.

\subsection{Preliminaries}\label{sct_r_etc}
In \cite{rathnayake2021observer, rathnayake2022sampled}, the authors propose observer-based R-CETC approaches for a class of reaction-diffusion PDEs with constant coefficients. The full-state feedback design for the case with a spatially varying reaction coefficient follows essentially the same steps. Therefore, we briefly present the results of the design below without going into the details of the derivations.

Events under the R-CETC are generated by the following dynamic event-trigger:
\begin{equation}\label{tnxt}
    t_{j+1}^r=\inf\big\{t\in\mathbb{R}_+\vert t>t_j^r,\Gamma^r(t)>0,j\in\mathbb{N}\big\},
\end{equation}
with $t^r_0=0$. The function $\Gamma^r(t)$ is given by 
\begin{equation}\label{gmmt}
  \Gamma^r(t)=d^2(t)-\gamma m^r(t),
\end{equation}
where $\gamma>0$ is an event-trigger design parameter. The function $d(t)$ satisfies \eqref{dt} for all $t\in[t_j^r,t^r_{j+1}),j\in\mathbb{N},$ and after the control input is updated at each $t=t_j^r$, $d(t_j^r)$ becomes zero by the virtue of \eqref{dt}. The dynamic variable $m^r(t)$ evolves according to the ODE
\begin{equation}\label{obetbc3m}\begin{split}
\dot{m}^r(t)=&-\eta m^r(t)-\rho d^{2}(t)+\beta_{1}\Vert u[t]\Vert^{2}+\beta_{2}u^2(1,t),\end{split}
\end{equation}
valid for all $t\in(t_{j}^r,t_{j+1}^r),j\in\mathbb{N}$ with $m^r(t_{0}^r)=m^r(0)>0$ and $m^r(t_{j}^{r-})=m^r(t^r_{j})=m^r(t_{j}^{r+})$, and $\eta,\rho,\beta_1,\beta_2>0$ being event-trigger parameters. 

Under Assumption \ref{ass1} with appropriately chosen event-trigger parameters (see Section \ref{ass2}), and by applying the event-triggered control input
\begin{equation}\label{etcl}
U_{j}^r=\int_{0}^{1}k(y)u(y,t_{j}^r)dy,
\end{equation}
in a zero-order hold manner between events generated by the event-trigger \eqref{tnxt}-\eqref{obetbc3m}, it can be shown that
\begin{itemize}
    \item The time between two consecutive events is positively and uniformly lower bounded by a constant $\tau>0$, \textit{i.e.,} $t^r_{j+1}-t^r_j\geq \tau,j\in\mathbb{N}$, where
    \begin{equation}\label{mdt}
\tau=\frac{1}{a}\ln \Big ( 1+\frac{\sigma a}{(1-\sigma)(a+\gamma\rho)}\Big),
\end{equation}
with $\sigma\in(0,1)$ and \begin{equation}\label{hhnm}
    a = 1+\rho_1+\eta>0.
\end{equation}
Here, $\rho_1$ is given by
\begin{equation}\label{aaappo}
    \rho_1=3\varepsilon^{2}k^{2}(1),
\end{equation}
where $k(y)$ is given by \eqref{ase1}.
\item The dynamic variable $m^r(t)$ governed by \eqref{obetbc3m} with $m^r(0)>0$, satisfies $m^r(t)>0$ for all $t>0$.

\item  A Lyapunov candidate defined as
\begin{equation}\label{apry}
    V^r(t) \triangleq \frac{B}{2}\Vert w[t]\Vert^2+m^r(t),
\end{equation}
for a suitably chosen $B>0$ (see Section \ref{ass2}) and with $w$ being the target system state governed by \eqref{tup}-\eqref{tup3}, satisfies
\begin{equation}\label{hhf}
    V^r(t)\leq e^{-b^*t}V_0,
\end{equation}
for all $t>0$, where $V_0=V^r(0)$ and for some $b^*>0$.

 \item The closed-loop solution of \eqref{ctpe1},\eqref{ctpe2},\eqref{etcla},\eqref{mctpe3x},\eqref{tnxt}-\eqref{etcl} globally exponentially converges to zero in the spatial   $L^2$ norm, satisfying the estimate
\begin{equation}\label{rrtx}
    \Vert u[t]\Vert\leq M e^{-\frac{b^{*}}{2}t}\sqrt{\Vert u[0]\Vert^2 +m^r(0)},
\end{equation}
for some $M,b^*>0$. Definitions of $M$ and $b^*$  are provided in Section 3.2. \hfill $\square$
\end{itemize}

We refer to the signal $e^{-b^*t}V_0$ in \eqref{hhf} as the \textit{performance-barrier} that should not be violated. By differentiating \eqref{apry} for $t\in (t_j^r,t_{j+1}^r),j\in\mathbb{N}$  along the solutions of \eqref{tup}-\eqref{tup3},\eqref{obetbc3m} and subject to Assumption \ref{ass1} along with appropriately chosen event-trigger parameters, one can obtain that
\begin{equation}\label{byu}
    \dot{V}^r(t)\leq -b^*V^r(t),
\end{equation}
for $t\in (t_j^r,t_{j+1}^r),j\in\mathbb{N}$. The relation \eqref{byu} indicates that the R-CETC forces the Lyapunov function \eqref{apry} to strictly decrease along the system trajectories. This stringent condition may limit our ability to harness the full potential of ETC for achieving sparse control updates. A more flexible approach might involve a triggering mechanism that permits the Lyapunov function to deviate from a monotonic decrease, yet remain compliant with the performance-barrier $e^{-b^*t}V_0$. Such flexibility could potentially result in longer intervals between events, \textit{i.e.,} an increase in dwell-times. In Section \ref{nnmd}, we introduce designs that embody this flexible approach.

\subsection{P-ETC}\label{nnmd}
We incorporate a \textit{performance residual}, defined as the difference between the value of the performance-barrier $e^{-b^*t}V_0$ and the Lyapunov function in the construction of the novel triggering mechanism. This inclusion is made with the intention of imparting greater flexibility to the behavior of the closed-loop system Lyapunov function, thereby permitting it to deviate from a monotonic decrease while adhering to the nominal performance. 

In Section \ref{ss1}, we introduce the P-CETC design, followed by its extensions to P-PETC and P-STC designs in Sections \ref{sct_P_PETC} and \ref{sct_P_STC}, respectively. These extensions aim to eliminate the need for continuous checking of the triggering function required in the P-CETC. Prior to detailing these designs, we state the following assumption about the parameters involved.

\subsubsection{Selection of Event-trigger Parameters}\label{ass2}
The designs to be introduced contain several parameters referred to as \textit{event-trigger parameters}: $\gamma,\eta,c,\beta_1,\beta_2,\rho$. Below, we present the conditions for appropriate choices of these parameters.

The parameters $\gamma,\eta,c>0$ are arbitrary design parameters, and $\beta_1,\beta_2>0$ are chosen as
\begin{equation}\label{betas}
\beta_{1}=\frac{\alpha_{1}}{\gamma(1-\sigma)},\hspace{5pt}\beta_{2}=\frac{\alpha_{2}}{\gamma(1-\sigma)},
\end{equation}
where $\sigma\in(0,1)$ and 
\begin{align}
\begin{split}
\label{al1}\alpha_{1}&=3\int_{0}^{1}\Big(\varepsilon k''(y)+\varepsilon k(1)k(y)+\lambda(y) k(y)\Big)^{2}dy,
\end{split}\\
\label{al2}
\alpha_{2}&=3\big(\varepsilon qk(1)+\varepsilon k'(1)\big)^{2},
\end{align}
with $k(y)$ given by \eqref{ase1}. Subject to Assumption \ref{ass1}, let a parameter $\rho>0$ be chosen as 
\begin{equation}\label{hhjknbv}
\rho=\frac{\varepsilon\kappa B}{2},
\end{equation}
where $B,\kappa>0$ are chosen such that
\begin{equation}\label{Bs}
\begin{split}
B\bigg(\varepsilon\min\Big\{\wp-\frac{\theta_1}{2},\frac{1}{2}\Big\}&-\frac{\varepsilon}{2\kappa}\bigg)-2\beta_1\tilde{L}^2-2\beta_{2}\\&-4\beta_{2}\int_{0}^1L^2(1,y)dy>0,
\end{split}
\end{equation}
with $\wp$ given by \eqref{rt}. Due to Assumption \ref{ass1}, it definitely holds that $\wp=q-\frac{1}{2\varepsilon}\int_{0}^{1}\lambda(y)dy>\theta_1/2$. In \eqref{Bs}, $\tilde{L}$ is 
\begin{equation}\label{tildel}
\tilde{L}=1+\Big(\int_{0}^{1}\int_{0}^xL^2(x,y)dydx\Big)^{1/2},
\end{equation}
with $L(x,y)$ satisfying \eqref{ctckeg}-\eqref{kkllspip}.

\subsubsection{P-CETC}\label{ss1}

Let $I^p =\{t^p_0,t^p_1,t^p_2,\ldots\}$ denote the sequence of event-times associated with the P-CETC\footnotemark. Let the event-trigger parameters $\eta,\gamma,c,\beta_1,\beta_2,\rho>0$ be selected as outlined in Section \ref{ass2}. The proposed P-CETC strategy consists of two components:

\footnotetext{A preliminary version of the P-CETC design appears in \cite{rathnayake2023pfmnc}.}

\begin{enumerate}
    \item An event-triggered boundary control input $U_j^p$
    \begin{equation}\label{obetbcp4a}
U_{j}^p=\int_{0}^{1}k(y)u(y,t_{j}^p)dy,
    \end{equation} valid for all $t\in[t_{j}^p,t_{j+1}^p),j\in\mathbb{N}$. Accordingly, the boundary condition \eqref{ctpe3} becomes
\begin{align}\label{ppp18}
    u_{x}(1,t)+qu(1,t)&=U_j^p.
\end{align}

    \item A dynamic event-trigger determining the event-times
    \begin{equation}\label{petzq}
    \begin{split}
    t_{j+1}^p=\inf\big\{t\in\mathbb{R}_+\vert& t>t_j^p,\Gamma^p(t)>0\big\},
    \end{split}
\end{equation}
with $t_0^p=0$ and $\Gamma^p(t)$ defined as  
\begin{equation}\label{mgw}
    \Gamma^p(t) \triangleq d^2(t)-\gamma m^p(t)-\frac{c}{\rho} W^p(t).
\end{equation}
Here $m^p(t)$ satisfies \begin{equation}\label{ggl}
\begin{split}
    \dot{m}^p(t)=&-\eta m^p(t)-\rho d^2(t)+\beta_1\Vert u[t]\Vert^2+\beta_2 u^2(1,t)\\&+c W^p(t),
\end{split}
\end{equation}
for $t\in(t_j^p,t_{j+1}^p),j\in\mathbb{N}$, where $m^p(t^p_0)=m^r(t^r_0)>0$, $m^p(t_j^{p-})=m^p(t_j^{p})=m^p(t^{p+}_{j})$, $d(t)$ is given by \eqref{dt} for $t\in[t_j^p,t_{j+1}^p),j\in\mathbb{N}$, and $W^p(t)$ is defined as
\begin{equation}\label{aacbnmh}
    W^p(t) \triangleq e^{-b^*t}V_0-V^p(t).
\end{equation}
In \eqref{aacbnmh}, $V^p(t)$ is defined as
\begin{equation}\label{abx}
    V^p(t)  \triangleq \frac{B}{2}\Vert w[t]\Vert^2+m^p(t),
\end{equation}
with
\begin{equation}\label{asf}
    V^p(0)=V^r(0)=V_0,
\end{equation}
$w$ satisfying the target PDE \eqref{tup}-\eqref{tup3} for $t\in[t_j^p,t_{j+1}^p),j\in\mathbb{N}$, and $B>0$ chosen to satisfy \eqref{Bs}. The exponential decay $b^*$ in \eqref{aacbnmh} is defined as  
\begin{equation}\label{BBB}
     b^* \triangleq\min\Big\{\frac{2b}{B},\eta\Big\}>0,
\end{equation}
with $b>0$ given by
\begin{equation}\label{aaaae}
     b=\frac{\varepsilon B}{4}-\beta_1\tilde{L}^2-2\beta_2\int_{0}^{1}L^2(1,y)dy.
 \end{equation}
 Here, $L(x,y)$ satisfies \eqref{ctckeg}-\eqref{kkllspip}, and $\tilde{L}$ is given by \eqref{tildel}. Note from \eqref{Bs} that $b>0$. 
\end{enumerate}

The term $W^p(t) \triangleq e^{-b^*t}V_0-V^p(t)$ is the so-called \textit{performance residual}, which is the difference between the value of the performance-barrier $e^{-b^*t}V_0$ and the Lyapunov function $V^p(t)$. The decay rate $b^*$ is the nominal exponential convergence rate guaranteed under the R-CETC briefed in Section \ref{sct_r_etc}. We introduce this residual into the triggering mechanism allowing the Lyapunov function to occasionally increase, provided it stays below the nominal performance-barrier $e^{-b^*t}V_0$. 

Next we present Lemmas \ref{qtpo} and \ref{auh} which are instrumental in proving the main results of the P-CETC that will be presented in Theorem \ref{exv}.


\begin{lemm}\label{qtpo}
    Under the P-CETC event-trigger \eqref{petzq}-\eqref{aaaae}, it holds that $d^{2}(t)\leq\gamma m^p(t)+\frac{c}{\rho}W^p(t)$, and consequently, $m^p(t)> 0,$ for all $t\in [0,\sup(I^p))$, where $I^p$ is the set of event-times $I^p=\{t_j^p\}_{j\in\mathbb{N}}$ with $t_j^p=0$.
\end{lemm}
\textbf{Proof} The P-CETC events are triggered to guarantee $\Gamma^p(t)\leq 0$, \textit{i.e.,} $d^{2}(t)\leq\gamma m^p(t)+\frac{c}{\rho}W^p(t)$ for $t\in [0,\sup(I^p))$. This inequality in combination with \eqref{ggl} yields:  
\begin{equation*}\label{lem1e1}
\begin{split}
\dot{m}^p(t)\geq&-(\eta+\gamma\rho) m^p(t)+\beta_{1}\Vert u[t]\Vert^{2}+\beta_{2}u^2(1,t)\\\geq&-(\eta+\gamma\rho) m^p(t),
\end{split}
\end{equation*}for $ t\in(t_{j}^p,t_{j+1}^p),j\in\mathbb{N}.$ Thus, considering the time-continuity of $m^p(t)$, we obtain the following estimate:
\begin{equation}\label{fa}
\begin{split}
&m^p(t)\geq m^p(t_{j}^p)e^{-(\eta+\gamma\rho)(t-t_{j}^p)},
\end{split}
\end{equation}
for $t\in[t_{j}^p,t_{j+1}^p],j\in\mathbb{N}$. Recall that we have chosen \mbox{$m^p(t_{0})=m^p(0)>0$}. Therefore, it follows from \eqref{fa} that $m^p(t)>0$  for all $t\in[0,t_{1}^p]$. Again using \eqref{fa} on $[t_{1}^p,t_{2}^p]$, we can show that $m^p(t)> 0$  for all $t\in[t_{1}^p,t_{2}^p]$. Applying the same reasoning successively to the future intervals, it can be shown that $m^p(t)> 0$ for $t\in [0,\sup(I^p))$. This completes the proof. \hfill $\square$

\begin{lemm}\label{auh}
Assume that an event has occurred at $t=t^*\geq 0$ under the P-CETC event-trigger \eqref{petzq}-\eqref{aaaae}. If the next event time $t=t^p$ generated by the P-CETC event-trigger is finite, then the next event time $t=t^r$ generated by the R-CETC event-trigger \eqref{tnxt}-\eqref{obetbc3m} is less than or equal to $t^p$, \textit{i.e.,} $t^r\leq t^p,$ provided that $m^r(t^*)=m^p(t^*)>0$ and $W^p(t)\geq 0$ for all $t\in[t^*,t^r]$. The equality holds if $W^p(t)=0$ for all $t\in [t^*,t^{r}=t^{p}]$.
\end{lemm}
\noindent\textbf{Proof} Consider the time period $t\in [t^*,\min\{t^r,t^p\}]$. Then, subtracting \eqref{obetbc3m} from \eqref{ggl} and assuming it is provided that $W^p(t)\geq 0$ for $t\in [t^*,\min\{t^{r},t^{p}\}]$, we obtain that 
\begin{equation}\label{wsq}
\begin{split}
    \dot{m}^p(t)-\dot{m}^r(t) &= -\eta\big(m^p(t)-m^r(t)\big)+cW^p(t)\\&\geq -\eta\big(m^p(t)-m^r(t)\big),
\end{split}
\end{equation}
for all $t\in (t^*,\min\{t^{r},t^{p}\})$. Here, the equality holds if $W^p(t)=0$ for all $t\in [t^*,\min\{t^{r},t^{p}\}]$. Solving \eqref{wsq} for $m^p(t)-m^r(t)$ in $t\in [t^*,\min\{t^r,t^p\}]$ and recalling the assumption that $m^p(t^*)=m^r(t^*)$, we obtain
\begin{equation*}
    m^p(t)-m^r(t) \geq e^{-\eta(t-t^*)}\big(m^p(t^*)-m^r(t^*)\big)=0,
\end{equation*} and therefore,
\begin{equation}\label{op1}
    m^p(t)\geq m^r(t), \forall t\in [t^*,\min\{t^{r},t^{p}\}].
\end{equation}
Here, the equality holds if $W^p(t)=0$ for all $t\in [t^*,\min\{t^{r},t^{p}\}]$.

Assume that $t^{r}>t^{p}$.
Then, we have from \eqref{op1} that 
\begin{equation}\label{pryu1}
    m^p(t)\geq m^r(t), \forall t\in [t^*,t^p],
\end{equation}
and from  \eqref{petzq},\eqref{mgw} that
\begin{equation}\label{pryu2}
d^2(t^p)=\gamma m^p(t^p)+\frac{c}{\rho}W^p(t^p),
\end{equation}
and from \eqref{tnxt},\eqref{gmmt} that
\begin{equation}\label{pryu3}
    d^2(t^p) < \gamma m^r(t^p).
\end{equation}
But \eqref{pryu1}-\eqref{pryu3} is a contradiction. Thus, 
$t^r\leq t^p$, with the equality being true if $W^p(t)=0$ for all $t\in [t^*,t^r=t^p]$. This completes the proof. \hfill $\square$


\begin{thme}\textnormal{\textbf{(Results under P-CETC)}}\label{exv}  
Consider the P-CETC approach \eqref{obetbcp4a}-\eqref{aaaae} under Assumption \ref{ass1}, which generates a set of event-times $I^p=\{t_j^p\}_{j\in\mathbb{N}}$ with $t_0^p=0$. It holds that
\begin{equation}
    \Gamma^p(t)\leq 0 \text{ for } \text{ all }t\in \big[0,\sup(I^p)\big).
\end{equation}
Consequently, if the event-trigger parameters $\eta,\gamma,c,\beta_1,\beta_2,\\\rho>0$ are chosen as outlined in Section \ref{ass2}, then the following results hold:
\begin{enumerate}
    \item[R1:] The set of event-times $I^p$ generates an increasing sequence. Specifically, it holds that $t^p_{j+1}-t^p_j\geq \tau>0$ where $\tau$ is given by \eqref{mdt}. Thus $t_j^p\rightarrow \infty$ as $j\rightarrow \infty$ excluding Zeno behavior.
 
\item [R2:] For every $u[0]\in L^{2}(0,1)$, there exists a unique solution $u:\mathbb{R}_+\times[0,1]\rightarrow\mathbb{R}$ such that $u\in C^0(\mathbb{R}_+;L^2(0,1)\cap C^1(J^p\times [0,1])$ with $u[t]\in C^2([0,1])$ which satisfy \eqref{ctpe2},\eqref{obetbcp4a},\eqref{ppp18} for all $t>0$ and \eqref{ctpe1} for all $t>0,x\in(0,1),$ where $J^p=\mathbb{R_{+}}\text{\textbackslash}I^p$.

\item[R3:] The dynamic variable $m^p(t)$ governed by \eqref{ggl}-\eqref{aaaae} with $m^p(0)=m^r(0)>0$ satisfies $m^p(t)>0$ for all $t>0$.

\item[R4:] As a result of R1-R3, the Lyapunov candidate $V^p(t)$ given by \eqref{abx},\eqref{asf} satisfies 
\begin{equation}\label{zzzbnmk}
    \dot{V}^p(t) \leq -b^* V^p(t)+c\big(e^{-b^*t}V_0-V^p(t)\big),
\end{equation}
for all $t\in (t_j^p,t_{j+1}^{p}),j\in\mathbb{N}$, and consequently,
\begin{equation}\label{asrt}
    V^p(t)\leq e^{-b^*t}V_0,
\end{equation}
for all $t>0$, where $b^*>0$ is given by \eqref{BBB}.

\item[R5:] The closed-loop solution of \eqref{ctpe1},\eqref{ctpe2},\eqref{obetbcp4a}-\eqref{aaaae} globally exponentially converges to zero in the spatial   $L^2$ norm satisfying \begin{equation}\label{rrt}
    \Vert u[t]\Vert\leq M e^{-\frac{b^{*}}{2}t}\sqrt{\Vert u[0]\Vert^2 +m^p(0)},
\end{equation}
where $b^*$ is given by \eqref{BBB} and
\begin{equation}\label{MMm}
    M=\sqrt{\frac{2\tilde{L}^2}{B}\max\Big\{\frac{B\tilde{K}^2}{2},1\Big\}}.
\end{equation}
 Here,  $\tilde{K}=1+\Big(\int_{0}^{1}\int_{0}^xK^2(x,y)dydx\Big)^{1/2}$ where $K(x,y)$ satisfies \eqref{ctcke1}-\eqref{aamlper}.
\end{enumerate}
\end{thme}

\begin{figure}
\centering
\includegraphics[scale=0.4]{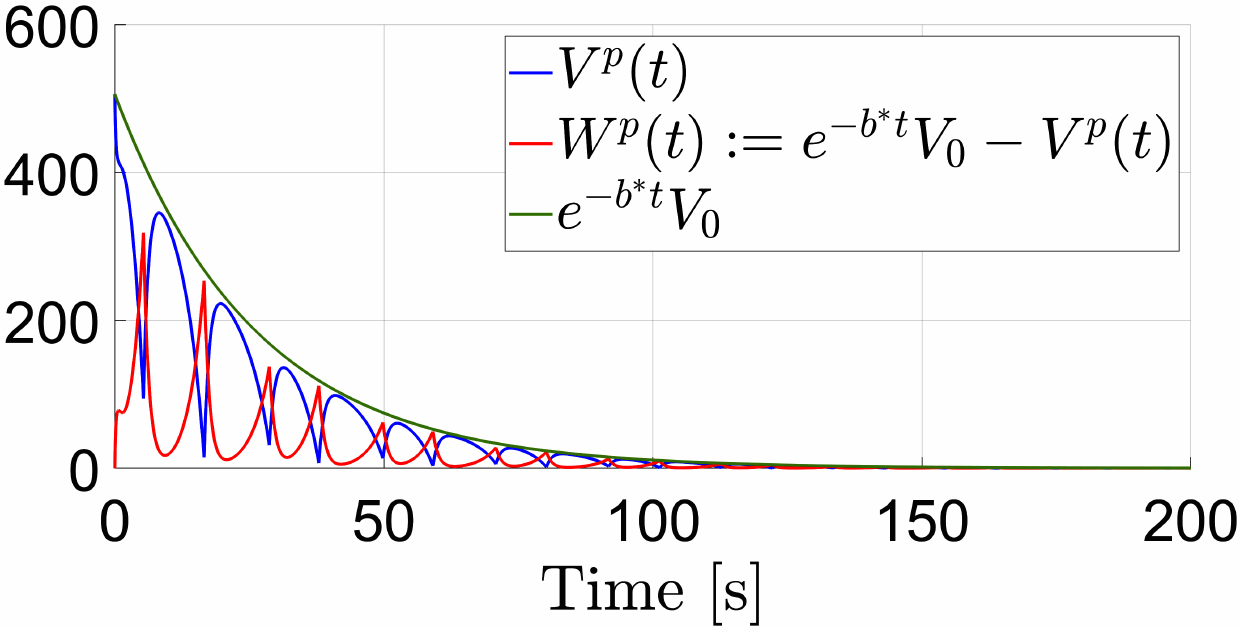}
\caption{Evolution of the Lyapunov function $V^p(t)$ and the performance residual $W^p(t)$.}
\label{acnw1}
\end{figure}

\noindent\textbf{Proof} Under the P-CETC, recall from Lemma \ref{qtpo} that it holds that $\Gamma^p(t)\leq 0$ and $m^p(t)> 0$ for $t\in[0,\sup(I^p))$. Consider the time period $t\in[0,\sup(I^p))$. By selecting the event-trigger parameters $\eta,\gamma,c,\beta_1,\beta_2,\rho>0$ as outlined in Section \ref{ass2} and using similar arguments provided in the proof of Theorem 2 in \cite{rathnayake2021observer}, we show that $V^p(t)$ given by \eqref{abx} satisfies 
\begin{equation}\label{abv}
    \dot{V}^p(t)\leq -b^* V^p(t)+cW^p(t),
\end{equation}
for $t\in(t_j^p,t_{j+1}^p),j\in\mathbb{N}$. Taking the time derivative of $W^p(t)$ given by \eqref{aacbnmh} and using \eqref{abv}, we show that
\begin{equation}\label{aqur}
\begin{split}
    \dot{W}^p(t) & = -b^* e^{-b^* t}V_0-\dot{V}^p(t)\\&\geq -(b^*+c)W^p(t),
\end{split}
\end{equation}
for $t\in (t_j^p,t_{j+1}^p),j\in\mathbb{N}$. Then, noting that $W^p(t)$ is continuous and $W^p(0)=0$, we obtain that
\begin{equation}\label{ccvv}
\begin{split}
    W^p(t)&\geq e^{-(b^*+c)(t-t_j^p)}W^p(t_j^p)\\
&\geq e^{-(b^*+c)(t-t_{j}^p)}\times
\prod_{i=1}^{i=j}e^{-(b^*+c)(t_{i}^p-t_{i-1}^p)} W^p(0)\\&\geq 
e^{-(b^*+c) t}W^p(0)=0.
\end{split}
\end{equation}
for all $t\in [0,\sup(I^p))$. Thus, recalling Lemma \ref{auh}, we state that $t^p_{j+1}-t^p_j\geq \tau$ where $\tau>0$ is the minimal-dwell time of the R-CETC as given by \eqref{mdt}, and $t_j^p\rightarrow \infty$ as $j\rightarrow \infty$ excluding Zeno behavior. The well-posedness of the controlled-plant \eqref{ctpe1},\eqref{ctpe2},\eqref{obetbcp4a},\eqref{ppp18} in the sense of R2 of Theorem \ref{exv} is a direct consequence of Proposition \ref{cor1}. The solution for all $t>0$ is constructed
iteratively between consecutive triggering times. Since the system is Zeno-free, we obtain that $m^p(t)$ governed by \eqref{ggl} with $m^p(0)>0$ satisfies $m^p(t)>0$, $V^p(t)$ satisfies \eqref{zzzbnmk} for all $t\in(t_j^p,t_{j+1}^p),j\in\mathbb{N}$, and $W^p(t)\geq 0$, \textit{i.e.,} $e^{-b^*t}V_0\geq V^p(t)$ for all $t>0$. Thus, following classical arguments involving the bounded invertibility of the transformations \eqref{ctbtd}-\eqref{aamlper} and \eqref{puoi}-\eqref{kkllspip}, we obtain the global exponential convergence of the solution of \eqref{ctpe1},\eqref{ctpe2},\eqref{obetbcp4a}-\eqref{aaaae} satisfying the decay estimate \eqref{rrt},\eqref{MMm}.  This completes the proof. 

\hfill $\square$ 

Unlike the R-CETC, which mandates the Lyapunov function to strictly decrease, the P-CETC (as well as P-PETC and P-STC, which will be discussed later) offers more flexibility to the Lyapunov function. The relation \eqref{zzzbnmk} implies that the time derivative of the Lyapunov function does not have to be negative at all times. Specifically, when $e^{-b^* t}V_0-V^p(t)$ is large, meaning the Lyapunov function is significantly below the performance-barrier, $\dot{V}^p(t)$ can be positive, allowing $V^p(t)$ to increase. Conversely, when $e^{-b^* t}V_0-V^p(t)$ is small, indicating the Lyapunov function is close to the performance-barrier, $\dot{V}^p(t)$ is compelled to decrease, even becoming negative. If $e^{-b^* t}V_0=V^p(t)$, then $\dot{V}^p(t)$ is definitely negative, preventing the Lyapunov function from breaching the performance-barrier and ensuring it remains below this threshold. Fig. \ref{acnw1} illustrates the evolution of $V^p(t)$ and the residual $W^p(t)=e^{-b^*t}V_0-V^p(t)$ in the simulation example considered in Section 4.

\begin{figure}
\centering
\includegraphics[scale=0.4]{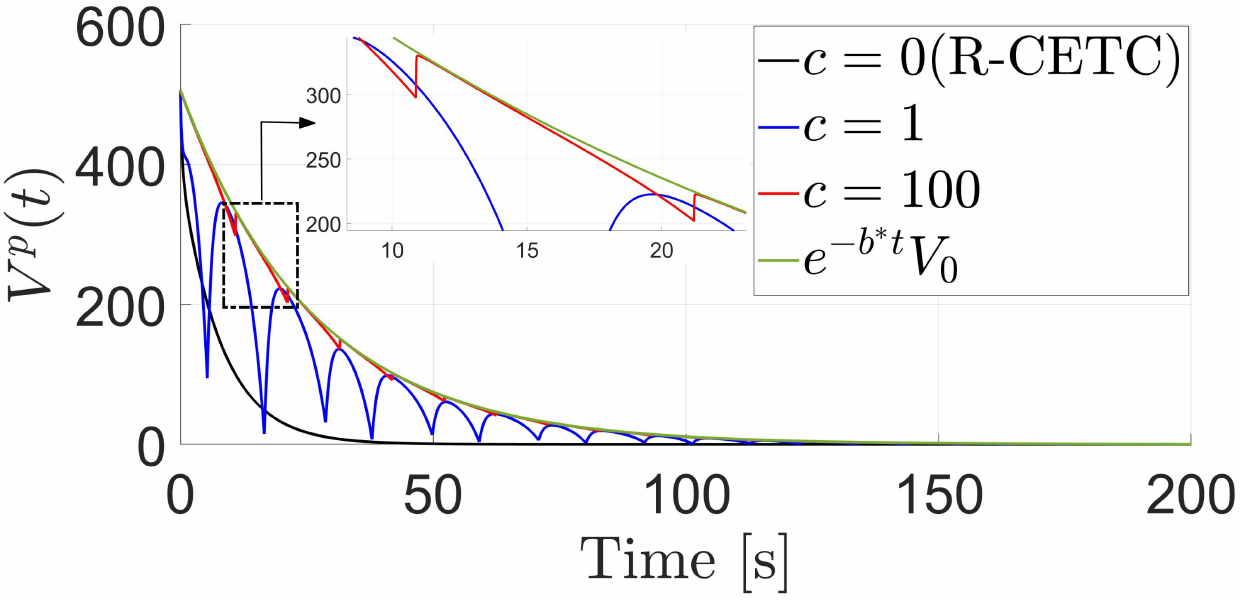}
\caption{Behavior of the Lyapunov function $V^p(t)$ under different choices of $c$. }
\label{acnw11}
\end{figure}

\begin{rmk}\label{sclpf}\rm
The parameter $c>0$ in the P-CETC event-trigger \eqref{petzq}-\eqref{aaaae} plays a pivotal role in shaping the behavior of the Lyapunov function $V^p(t)$ given by \eqref{abx}. For any $c$ such that $0<c<\infty$, it follows from R4 of Theorem \ref{exv} that $\dot{V}^p(t)\leq -b^* V^p(t)+c\big(e^{-b^* t}V_0-V^p(t)\big),$ for all $t\in(t_j^p,t_{j+1}^p),j\in\mathbb{N}$, and consequently, $V^p(t)\leq e^{-b^*t}V_0$, for all $t>0$. As described earlier, this allows the Lyapunov function to increase during certain periods as long as it remains below the nominal performance-barrier. As $c\rightarrow \infty$, the Lyapunov function approaches the performance-barrier, \textit{i.e.,} $V^p(t)\rightarrow e^{-b^*t}V_0$. Conversely, setting $c=0$ reduces the P-CETC to the R-CETC, resulting in $\dot{V}^p(t)\leq -b^* V^p$(t). This demands a strict decrease in the Lyapunov function while ensuring $V^p(t)\leq e^{-b^*t}V_0$ for all $t>0$. Fig. \ref{acnw11} illustrates the behavior of the Lyapunov function for different choices of $c$ in the simulation example considered in Section 4. 
\end{rmk}

\begin{rmk}\rm Recall from R4 of Theorem \ref{exv} that $V^p(t)\leq e^{-b^*t}V_0$, for all $t>0$. Therefore, it follows from Lemma \ref{auh} that the time of the next event generated by the P-CETC approach \eqref{petzq}-\eqref{aaaae} is greater than or equal to the time of the next event generated by the R-CETC approach \eqref{tnxt}–\eqref{obetbc3m}, with the equality occurring if $V^p(t) = e^{-b^*t}V_0$ for all times between events. If $V^p(t)$ deviates from being equal to the barrier $e^{-b^*t}V_0$ for any period between two events, the time of the next event generated by the P-CETC approach will be strictly greater than the time of the next event generated by the R-CETC approach. As mentioned in Remark \ref{sclpf}, when $c \rightarrow \infty$, we have that $V^p(t) \rightarrow e^{-b^\star t}V_0$ for all $t>0$ (see Fig. \ref{acnw11} also). For a parameter $c$ chosen such that $0 < c < \infty$, it is unlikely that $V^p(t) = e^{-b^*t}V_0$ for all times between events, thereby almost always resulting in an event time strictly greater than that generated by the R-CETC.
\end{rmk}

\begin{rmk}\rm 
Due to R4 of Theorem \ref{exv}, in the absence of disturbances, the violation of the performance barrier by the Lyapunov function is not possible. However, in the presence of disturbances, the performance barrier may be violated. Particularly, when time reaches large values, the performance residual \( W^p(t) = e^{-b^\star t} V_0 - V^p(t) \) becomes more susceptible to disturbances. To guard the P-CETC design against disturbances, the following modification to the triggering mechanism can be made:
   \begin{equation}\label{mgwd}
    \Gamma^p(t) \triangleq d^2(t)-\gamma m^p(t)-\frac{c}{\rho} \max\big\{0,W^p(t)\big\},
\end{equation}
with $m^p(t)$ satisfying \begin{equation}\label{ggld}
\begin{split}
    \dot{m}^p(t)=&-\eta m^p(t)-\rho d^2(t)+\beta_1\Vert u[t]\Vert^2+\beta_2 u^2(1,t)\\&+c \max\big\{0,W^p(t)\big\},
\end{split}
\end{equation}
for $t\in(t_j^p,t_{j+1}^p),j\in\mathbb{N}$, and $m^p(t^p_0)=m^r(t^r_0)>0$, $m^p(t_j^{p-})=m^p(t_j^{p})=m^p(t^{p+}_{j})$.\\ \indent With this modification, if the Lyapunov function violates the performance barrier due to a disturbance (i.e., \(W^p(t) < 0\)), P-CETC hands over the operation to R-CETC. This should be seen not as a drawback, but rather as a safety feature of the design. When disturbances occur, sparse control updates are unfavorable, and frequent control updates are advised to minimize the time the plant operates in open loop. The above modification achieves this by handing over the operation to R-CETC whenever a disturbance large enough to violate the performance barrier occurs, as R-CETC generates more frequent control updates compared to P-CETC. Since R-CETC forces the Lyapunov function to strictly decrease (see \eqref{byu}), it is even possible for the Lyapunov function to fall below the performance barrier again, leading to the operation being handed back to P-CETC.
\end{rmk}

\subsubsection{P-PETC}\label{sct_P_PETC}
In this subsection, we present the P-PETC approach derived from the P-CETC scheme \eqref{obetbcp4a}-\eqref{aaaae}, redesigning the triggering function $\Gamma^p(t)$ from \eqref{mgw} to $\tilde{\Gamma}^p(t)$ for periodic evaluation. This adjustment, coupled with event-triggered control updates, ensures $\Gamma^p(t)$ remains non-positive and $m^p(t)$ remains positive along the P-PETC closed-loop system solution. As a result, convergence properties equivalent to \textit{R4} and \textit{R5} in Theorem \ref{exv} holds for the P-PETC closed-loop system as well. 

Let $\tilde{I}^p =\{\tilde{t}^p_0,\tilde{t}^p_1,\tilde{t}^p_2,\ldots\}$ denote the sequence of event-times associated with the P-PETC. Let the parameters $\gamma,c,\beta_1,\beta_2,\rho>0$ be selected as outlined in Section \ref{ass2}, and let $\eta>0$ be selected later. The proposed P-PETC strategy consists of two components:

\begin{enumerate}
    \item An event-triggered boundary control input $\tilde{U}_j^p$
    \begin{equation}\label{obetbcp4f}
\tilde{U}^p_{j}=\int_{0}^{1}k(y)u(y,\tilde{t}^p_{j})dy,
    \end{equation} valid for all $t\in[\tilde{t}^p_{j},\tilde{t}^p_{j+1}),j\in\mathbb{N}$. Accordingly, the boundary condition \eqref{ctpe3} becomes
\begin{align}\label{ppp1f}
    u_{x}(1,t)+qu(1,t)&=\tilde{U}^p_j.
\end{align}

    \item A periodic event-trigger determining the event-times
    \begin{equation}\label{petwq}
    \begin{split}
    \tilde{t}^p_{j+1}=\inf\big\{t\in\mathbb{R}_+\vert& t>\tilde{t}^p_j,\tilde{\Gamma}^p(t)>0,\\& \qquad t=nh, h>0, n\in\mathbb{N}\big\},
    \end{split}
\end{equation}
with $\tilde{t}^p_0=0$. Here, $h$ is the sampling period satisfying 
\begin{equation}\label{hjgfs}
    0< h\leq \tau,
\end{equation}
with $\tau$  given by \eqref{mdt}, and $\tilde{\Gamma}^p(t)$  defined as
\begin{equation}\label{tildeGqwrefa}
\begin{split}
\tilde{\Gamma}^p(t )=&
(a+\gamma\rho)e^{ah}d^{2}(t )-\gamma\rho d^2(t )-\gamma a m^p(t)\\&-\frac{ac}{\rho}e^{-ch}W^p(t).
\end{split}
\end{equation}
Here, $a$ is given by \eqref{hhnm}, $d(t)$ satisfies \eqref{dt} along the solution of \eqref{ctpe1},\eqref{ctpe2}, \eqref{obetbcp4f},\eqref{ppp1f} for all $t\in[\tilde{t}_j^p,\tilde{t}^p_{j+1}),j\in\mathbb{N},$ and $m^p(t)$ is governed by the ODE \eqref{ggl}-\eqref{aaaae} along the solution of  \eqref{ctpe1},\eqref{ctpe2},\eqref{obetbcp4f},\eqref{ppp1f} for $t\in(\tilde{t}_j^p,\tilde{t}_{j+1}^p),j\in\mathbb{N}$. 
\end{enumerate}


Next we present Lemmas \ref{lem2} and \ref{aqrt} and Proposition \ref{imptnt} which are instrumental in proving the main results of the P-PETC presented in Theorem \ref{cvgh}.

\begin{lemm}\label{lem2} Consider the P-PETC approach \eqref{obetbcp4f}-\eqref{tildeGqwrefa} which generates an increasing set of event-times $\tilde{I}^p=\{\tilde{t}_j^p\}_{j\in\mathbb{N}}$ with $\tilde{t}^p_0=0$.  For $d(t)$ given by \eqref{dt}, it holds that\begin{equation}\label{ghm}
\big(\dot{d}(t)\big)^2\leq \rho_{1} d^{2}(t)+\alpha_{1}\Vert u[t]\Vert^{2}+\alpha_{2}\vert u(1,t)\vert^{2},
\end{equation}
along the solution of  \eqref{ctpe1},\eqref{ctpe2},\eqref{obetbcp4f},\eqref{ppp1f} for all $t\in\big(nh,(n+1)h\big)$ and any $n\in\big[\tilde{t}^p_j/h,\tilde{t}^p_{j+1}/h\big)\cap\mathbb{N}$. Here, $\rho_1,\alpha_1$ and $\alpha_2$ are give by \eqref{aaappo},\eqref{al1}, and \eqref{al2}, respectively.\end{lemm}

The proof is very similar to that of Lemma 2 in \cite{rathnayake2021observer}, and hence omitted.


\begin{lemm}\label{aqrt}
Consider the P-PETC approach \eqref{obetbcp4f}-\eqref{tildeGqwrefa} under Assumption \ref{ass1}, which generates an increasing set of event-times $\tilde{I}^p=\{\tilde{t}_j^p\}_{j\in\mathbb{N}}$ with $\tilde{t}^p_0=0$. If the event-trigger parameters $\gamma,c,\beta_1,\beta_2,\rho>0$ are chosen as outlined in Section \ref{ass2}, and $\eta>0$ is chosen such that
\begin{equation}\label{qptya}
    \eta\leq \frac{2b}{B},\end{equation}
with $b>0$ given by \eqref{aaaae} and $B$ satisfying \eqref{Bs}, then for the residual $W^p(t)$ given by \eqref{aacbnmh}, the followings hold along the solution of  \eqref{ctpe1},\eqref{ctpe2},\eqref{tup}-\eqref{tup3},\eqref{ggl}-\eqref{aaaae},\eqref{obetbcp4f}-\eqref{tildeGqwrefa}:
 \begin{equation}\label{akp}
    W^p(t)\geq e^{-(b^*+c)(t-nh)}W^p(nh)\text{ with }b^*=\eta,
\end{equation}
 for all $t\in\big[nh,(n+1)h\big)$ and any $n\in\big[\tilde{t}^p_j/h,\tilde{t}^p_{j+1}/h\big)\cap\mathbb{N}$ and
\begin{equation}\label{akpq}
   W^p(t)\geq 0,\text{ i.e., } V^p(t)\leq e^{-b^*t}V_0\text{ with }b^*=\eta,
\end{equation}
for all $t>0$. 
\end{lemm}
\noindent\textbf{Proof} By selecting the event-trigger parameters $\gamma,c,\beta_1,\beta_2,\rho>0$ as outlined in Section \ref{ass2} with $\eta>0$ and differentiating \eqref{abx} along the solutions of \eqref{tup}-\eqref{tup3}, \eqref{ggl}-\eqref{aaaae} in $t\in(nh,(n+1)h)$ and $n\in\big[\tilde{t}^p_j/h,\tilde{t}^p_{j+1}/h\big)\cap\mathbb{N}$, one can obtain that
\begin{equation}\label{auri}
    \dot{V}^{p}(t)\leq -\Big(\frac{2b}{B}\Big)\frac{B}{2}\Vert w[t]\Vert^2-\eta m^p(t)+cW^p(t).
\end{equation}
If $m^p(t)>0$, then, we could have obtained the estimate \eqref{abv} from \eqref{auri}. However, it is still not known that $m^p(t)$ would remain positive along the P-PETC solution \eqref{ctpe1},\eqref{ctpe2},\eqref{ggl}-\eqref{aaaae},\eqref{obetbcp4f}-\eqref{tildeGqwrefa}.  Yet, if $\eta$ is selected as in \eqref{qptya}, we obtain from \eqref{auri} that $\dot{V}^p(t)\leq -\eta \Big(\frac{B}{2}\Vert w[t]\Vert^2+m^p(t)\Big)+cW^p(t)-\frac{B}{2}\Big(\frac{2b}{B}-\eta\Big)\Vert w[t]\Vert^2\leq 
    -b^* V^p(t)+c W^p(t).$ Note above that $b^*=\eta$ since $ b^*=\min\Big\{\frac{2b}{B},\eta\Big\}$ as given by \eqref{BBB} and $\eta\leq \frac{2b}{B}$. Then, following similar arguments as in \eqref{abv}-\eqref{ccvv}, we obtain the estimate \eqref{akp} valid for all $t\in\big[nh,(n+1)h\big)$ and any $n\in\big[\tilde{t}^p_j/h,\tilde{t}^p_{j+1}/h\big)\cap\mathbb{N}$. Similarly, we derive \eqref{akpq} which is valid for all $t>0$ owing to the event-times under the P-PETC \eqref{obetbcp4f}-\eqref{tildeGqwrefa} forming an increasing sequence. We omit the details of this derivation to avoid repetition. \hfill $\square$


Although Lemma \ref{aqrt} establishes that $V^p(t)\leq e^{-b^*t}V_0$ for any $t>0$ along the  P-PETC solution \eqref{ctpe1},\eqref{ctpe2},\eqref{ggl}-\eqref{aaaae},\eqref{obetbcp4f}-\eqref{tildeGqwrefa} with the event-trigger parameters $\gamma,c,\beta_1,\beta_2,\rho>0$ chosen as outlined in Section \ref{ass2}, and $\eta>0$ chosen to satisfy \eqref{qptya}, we are not yet in a position to assert the global exponential convergence properties \eqref{rrt},\eqref{MMm}. The reason is that the positivity of $m^p(t)$ along the  P-PETC solution \eqref{ctpe1},\eqref{ctpe2},\eqref{ggl}-\eqref{aaaae},\eqref{obetbcp4f}-\eqref{tildeGqwrefa} has not been proven, which is essential for ensuring the positive definiteness of $V^p(t)$ given by \eqref{abx}.


\begin{prop}\label{imptnt}  Consider the P-PETC approach \eqref{obetbcp4f}-\eqref{tildeGqwrefa} under Assumption \ref{ass1}, which generates an increasing set of event-times $\{\tilde{t}_j^p\}_{j\in\mathbb{N}}$ with $\tilde{t}^p_0=0$. If the event-trigger parameters $\gamma,c,\beta_1,\beta_2,\rho>0$ are chosen as outlined in Section \ref{ass2}, and $\eta>0$ is selected as in \eqref{qptya}, then $\Gamma^p(t)$ given by \eqref{mgw}  satisfies \begin{equation}\label{ineq}
 \begin{split}
    &\Gamma^p(t)\\&\leq \frac{1}{a}\Big((a+\gamma\rho)d^{2}(nh)e^{a(t-nh)}-\gamma\rho d^2(nh)\\&\quad\qquad-\gamma a m^p(nh)-\frac{ac}{\rho}e^{-c(t-nh)}W^p(nh)\Big)e^{-\eta(t-nh)},
\end{split}
\end{equation}
where $a$ is given by \eqref{hhnm} and $h$ is the sampling period chosen as in \eqref{hjgfs}, along the solution of  \eqref{ctpe1},\eqref{ctpe2},\eqref{ggl}-\eqref{aaaae},\eqref{obetbcp4f}-\eqref{tildeGqwrefa} for all $t\in\big[nh,(n+1)h\big)$ and any $n\in\big[\tilde{t}^p_j/h,\tilde{t}^p_{j+1}/h\big)\cap\mathbb{N}$. 
\end{prop}
\noindent \textbf{Proof}. Given the event-trigger parameters $\gamma,c,\beta_1,\beta_2,\rho>0$ are chosen as outlined in Section \ref{ass2} and  setting $\eta>0$ as specified in \eqref{qptya}, it holds that, $  W^p(t)\geq e^{-(b^*+c)(t-nh)}W^p(nh)$ with $b^*=\eta,$ for all $t\in\big[nh,(n+1)h\big)$ and any $n\in\big[\tilde{t}^p_j/h,\tilde{t}^p_{j+1}/h\big)\cap\mathbb{N}$, as derived in Lemma \ref{aqrt}. Thus, from \eqref{mgw}, it follows that  
\begin{equation}\label{apoi}
    \Gamma^p(t)\leq d^2(t)-\gamma m^{p}(t)-\frac{c}{\rho}e^{-(\eta+c)(t-nh)}W^p(nh),
\end{equation}
for all $t\in\big[nh,(n+1)h\big)$ and any $n\in\big[\tilde{t}^p_j/h,\tilde{t}^p_{j+1}/h\big)\cap\mathbb{N}$, along the solution of \eqref{ctpe1},\eqref{ctpe2},\eqref{ggl}-\eqref{aaaae},\eqref{obetbcp4f}-\eqref{tildeGqwrefa}. Let us define
\begin{equation}\label{qxb}
    \Gamma^{p*}(t) \triangleq d^2(t)-\gamma m^{p}(t).
\end{equation}
Taking the time derivative of \eqref{qxb} in $t\in(nh,(n+1)h)$ and $n\in\big[\tilde{t}^p_j/h,\tilde{t}^p_{j+1}/h\big)\cap\mathbb{N}$, using Young's inequality, the relation  \eqref{ghm}, and the dynamics of $m^p(t)$ given by \eqref{ggl}, we get
\begin{equation}\label{nnt}
    \begin{split}
        &\dot{\Gamma}^{p*}(t )=2d(t )\dot{d}(t )-\gamma \dot{m}^p(t)\\
        &\leq d^2(t )+\big(\dot{d}(t )\big)^2-\gamma \dot{m}^p(t)\\
        &\leq \big(1+\rho_1+\gamma\rho\big)d^2(t )+\gamma\eta m^p(t)-(\gamma\beta_1-\alpha_1)\Vert u[t]\Vert^2\\&\quad-(\gamma\beta_2-\alpha_2) u^2(1,t)-\gamma cW^p(t),
    \end{split}
\end{equation}
By using \eqref{qxb} to substitute $d^2(t)$ into \eqref{nnt}, we arrive at the following estimate:
\begin{equation}\label{hhm}
\begin{split}
    &\dot{\Gamma}^{p*}(t)\leq  (1+\rho_1+\gamma\rho)\Gamma^{p*}(t)+\gamma(a+\gamma\rho)m^p(t)\\&-(\gamma\beta_1-\alpha_1)\Vert u[t]\Vert^2-(\gamma\beta_2-\alpha_2) u^2(1,t)-\gamma cW^p(t),
\end{split}
\end{equation}
where $a$ is given by \eqref{hhnm}. The closed-loop system \eqref{ctpe1},\eqref{ctpe2},\eqref{obetbcp4f}-\eqref{tildeGqwrefa} has a unique solution for $t\in[\tilde{t}_j^p,\tilde{t}_{j+1}^p],j\in\mathbb{N}$ as per Proposition \ref{cor1}, and the system \eqref{ggl}-\eqref{aaaae} has a solution with $m^p(t)\in C^0([\tilde{t}^p_j,\tilde{t}^p_{j+1}];\mathbb{R})$. Since both sides of \eqref{hhm} are well-behaved in $t\in(nh,(n+1)h)$ and $n\in\big[\tilde{t}^p_j/h,\tilde{t}^p_{j+1}/h\big)\cap\mathbb{N}$, we can assert the existence of a non-negative function $\iota(t )\in C^0\big((\tilde{t}^p_{j},\tilde{t}^p_{j+1});\mathbb{R}_{+}\big)$ such that
\begin{equation}\label{rrghk}
\begin{split}
    \dot{\Gamma}^{p*}(t )=&(1+\rho_1+\gamma\rho)\Gamma^{p*}(t )+\gamma(a+\gamma\rho)m^p(t )\\&-(\gamma\beta_1-\alpha_1)\Vert u[t]\Vert^2-(\gamma\beta_2-\alpha_2) u^2(1,t)\\&-\gamma cW^p(t)-\iota(t ),
\end{split}
\end{equation}
for all $t\in(nh,(n+1)h)$ and $n\in\big[\tilde{t}^p_j/h,\tilde{t}^p_{j+1}/h\big)\cap\mathbb{N}$. 
Additionally, by using \eqref{qxb} to substitute $d^2(t)$ into \eqref{ggl}, we can rewrite the dynamics of $m^p(t)$ as follows:
\begin{equation}\label{hmm1}
\begin{split}
    \dot{m}^p(t) =& -\rho \Gamma^{p*}(t )-(\gamma\rho+\eta)m^p(t)+\beta_1\Vert u[t]\Vert^2\\&+\beta_2 u^2(1,t)+cW^p(t),
\end{split}
\end{equation}
for $t\in(nh,(n+1)h)$ and $n\in\big[\tilde{t}^p_j/h,\tilde{t}^p_{j+1}/h\big)\cap\mathbb{N}$. Then, combining \eqref{rrghk} with \eqref{hmm1}, we obtain the following ODE system
\begin{equation}\label{zv}
    \dot{z}(t )=Az(t )+v(t ),
\end{equation}
where
\begin{equation*}
\begin{split}
&z(t )=
\begin{bmatrix}
\Gamma^{p*}(t )\\m^p(t)
\end{bmatrix},
\text { }
    A = \begin{bmatrix}
1+\rho_1+\gamma\rho &  \gamma\big(a+\gamma\rho\big)\\
    -\rho & -(\gamma\rho+\eta)
    \end{bmatrix},\\&
    v(t )  =\begin{bmatrix}
    \Big(
    \begin{split}
        -(\gamma\beta_1-\alpha_1)\Vert u[t]\Vert^2-(&\gamma\beta_2-\alpha_2)u^2(1,t)\\&-\gamma cW^p(t)-\iota(t )
        \end{split}\Big)\\\beta_1\Vert u[t]\Vert^2+\beta_2u^2(1,t)+cW^p(t)
    \end{bmatrix}.
\end{split}
\end{equation*}
From \eqref{zv}, we obtain
\begin{equation*}
    z(t )= e^{A(t-nh)}z(nh )+\int_{nh}^{t} e^{A(t-\xi)}v(\xi )d\xi,
\end{equation*}
for all $t\in[nh,(n+1)h)$ and $n\in\big[\tilde{t}^p_j/h,\tilde{t}^p_{j+1}/h\big)\cap\mathbb{N}$, using which we show
\begin{equation}\label{rrto}
    \Gamma^{p*}(t )= Ce^{A(t-nh)}z(nh )+\int_{nh}^{t} Ce^{A(t-\xi)}v(\xi )d\xi,  
\end{equation}
where
\begin{equation*}
    C=\begin{bmatrix}
        1&&0
    \end{bmatrix}.
\end{equation*}
The matrix $A$ has two distinct eigenvalues $-\eta$ and $1+\rho_1$. Therefore, using matrix diagonalization, we express the matrix exponential $e^{At}$ as 
\begin{equation*}\label{expm}
\begin{split}
    e^{At}=\frac{\rho}{a}\begin{bmatrix}
        -\gamma & -\frac{a+\gamma\rho}{\rho}\\1&1
    \end{bmatrix}&\begin{bmatrix}
        e^{-\eta t}&0\\0& e^{\big(1+\rho_1\big)t}
    \end{bmatrix}\begin{bmatrix}
        1 & \frac{a+\gamma\rho}{\rho}\\-1&-\gamma
    \end{bmatrix}.
\end{split}
\end{equation*}
Moreover, it can be shown that 
\begin{equation}\label{mmh1}
\begin{split}
    &Ce^{A(t-\xi)}v(\xi )\\&=-\Big(\big(\gamma\beta_1-\alpha_1\big)g_1(t-\xi)-\beta_1 g_2(t-\xi)\Big)\Vert u[\xi]\Vert^2\\&\quad-\Big(\big(\gamma\beta_2-\alpha_2\big)g_1(t-\xi)-\beta_2 g_2(t-\xi)\Big)u^2(1,\xi)\\& \quad-c\Big(\gamma  g_1(t-\xi)-g_2(t-\xi)\Big)W^p(\xi)
    \\&\quad-g_1(t-\xi)\iota(\xi ),
\end{split}
\end{equation}
where
\begin{equation*}\label{g1}
    g_1(t)=\frac{1}{a}\Big(-\gamma\rho +(a+\gamma\rho)e^{at}\Big)e^{-\eta t},
\end{equation*}
\begin{equation*}\label{g2}
      g_2(t)=\frac{\gamma (a+\gamma\rho)}{a}\Big(-1+ e^{at}\Big)e^{-\eta t}.
\end{equation*}
We can easily observe that $g_1(t)>0$ for all $t\geq 0$. We also obtain that
\begin{equation*}
    \gamma g_1(t)- g_2(t) = \gamma e^{-\eta t}>0,
\end{equation*}
for all $t\geq 0$. Furthermore, noting that $\gamma\beta_i/\alpha_i=1/(1-\sigma), i=1,2$ from \eqref{betas}, and recalling \eqref{mdt}, we show that
\begin{equation}\label{nnj}
\begin{split}
 &\big(\gamma\beta_i-\alpha_i\big)g_1(t-\xi)-\beta_i g_2(t-\xi)\\&= \frac{\alpha_i (a+\gamma\rho) }{a}\bigg(1+\frac{\sigma a}{(1-\sigma)(a+\gamma\rho)}-e^{a(t-\xi)}\bigg)e^{-\eta (t-\xi)}\\&=
 \frac{\alpha_i (a+\gamma\rho) }{a}\bigg(e^{a\tau}-e^{a(t-\xi)}\bigg)e^{-\eta (t-\xi)},
\end{split}
\end{equation}
for $i=1,2$. As $nh\leq\xi\leq t<(n+1)h$, and $h\leq \tau$, we note from \eqref{nnj} that $\big(\gamma\beta_i-\alpha_i\big)g_1(t-\xi)-\beta_i g_2(t-\xi)>0$ for $i=1,2$. Thus, from \eqref{mmh1} along with the fact that $W^p(t)\geq 0$ from Lemma \ref{aqrt}, we are certain that $Ce^{A(t-\xi)}v(\xi )\leq 0$ for all $t,\xi$ such that $nh\leq\xi\leq t<(n+1)h$, and $n\in\big[\tilde{t}^p_j/h,\tilde{t}^p_{j+1}/h\big)\cap\mathbb{N}$. Considering this fact along with \eqref{rrto}, the following inequalities can be derived for $t\in[nh,(n+1)h)$
\begin{equation}\label{gghj}
    \begin{split}
        &\Gamma^{p*}(t )\leq Ce^{A(t-nh)}z(nh )\\
&\leq      g_1(t-nh)\Gamma^{p*}(nh )+g_2(t-nh)m^p(nh)\\&\leq 
\frac{1}{a}\Big(-\gamma(a+\gamma\rho)m^p(nh)-\gamma\rho\Gamma^{p*}(nh )\\&+(a+\gamma\rho)\big(\Gamma^{p*}(nh )+\gamma m^p(nh)\big)e^{a(t-nh)}\Big)e^{-\eta(t-nh)}.
    \end{split}
\end{equation}
By substituting $\Gamma^{p*}(nh )$ using \eqref{qxb} into \eqref{gghj} and recalling \eqref{apoi}, we obtain the inequality \eqref{ineq} that is valid for $t\in[nh,(n+1)h)$ completing the proof. \hfill$\square$


\begin{thme}\textnormal{\textbf{(Results under P-PETC)}}\label{cvgh}
 Consider the P-PETC \eqref{obetbcp4f}-\eqref{tildeGqwrefa} under Assumption \ref{ass1}, which generates an increasing set of event-times $\tilde{I}^p=\{\tilde{t}_j^p\}_{j\in\mathbb{N}}$ with $\tilde{t}_0^p=0$. For every $u[0]\in L^{2}(0,1)$, there exists a unique solution $u:\mathbb{R}_+\times[0,1]\rightarrow\mathbb{R}$ such that $u\in C^0(\mathbb{R}_+;L^2(0,1)\cap C^1(\tilde{J}^p\times [0,1])$ with $u[t]\in C^2([0,1])$ which satisfies \eqref{ctpe2},\eqref{obetbcp4f},\eqref{ppp1f} for all $t>0$ and \eqref{ctpe1} for all $t>0,x\in(0,1),$ where $\tilde{J}^p=\mathbb{R_{+}}\text{\textbackslash}\tilde{I}^p$. If the event-trigger parameters $\gamma,c,\beta_1,\beta_2,\rho>0$ are chosen as outlined in Section \ref{ass2}, and $\eta>0$ is selected as in \eqref{qptya}, then the following results hold:
\begin{enumerate}
    \item[R1:] $\Gamma^p(t)$ given by \eqref{mgw}  satisfies $\Gamma^p(t)\leq 0$ along the solution of \eqref{ctpe1},\eqref{ctpe2},\eqref{ggl}-\eqref{aaaae},\eqref{obetbcp4f}-\eqref{tildeGqwrefa} for all $t>0$.  

\item[R2:] The dynamic variable $m^p(t)$ governed by \eqref{ggl}-\eqref{aaaae} with $m^p(0)=m^r(0)>0$ satisfies $m^p(t)>0$ along the solution of \eqref{ctpe1},\eqref{ctpe2},\eqref{obetbcp4f}-\eqref{tildeGqwrefa} for all $t>0$.

\item[R3:] The Lyapunov candidate $V^p(t)$ given by \eqref{abx},\eqref{asf} satisfies \eqref{zzzbnmk} all $t\in (\tilde{t}_j^p,\tilde{t}_{j+1}^{p}),j\in\mathbb{N}$ and \eqref{asrt} for all $t>0$, with $b^*=\eta$ . 

\item[R4:] The closed-loop solution of \eqref{ctpe1},\eqref{ctpe2},\eqref{obetbcp4f}-\eqref{tildeGqwrefa} globally exponentially converges to zero in the spatial   $L^2$ norm satisfying the estimate \eqref{rrt},\eqref{MMm} with $b^*=\eta$.
\end{enumerate}
\end{thme}
\noindent\textbf{Proof} 
The well-posedness of the closed-loop system \eqref{ctpe1}, \eqref{ctpe2}, \eqref{obetbcp4f},\eqref{ppp1f} directly follows from Proposition \ref{cor1}. The system's solution for all $t>0$ is constructed iteratively between consecutive events. Given that the parameters $\gamma, c, \beta_1, \beta_2, \rho$ are chosen in accordance with Section \ref{ass2}, and with $\eta>0$ selected as indicated in \eqref{qptya}, Lemma \ref{aqrt} ensures that $W^p(t)\geq 0$ for all $t>0$. The subsequent analysis focuses on the dynamics of $\Gamma^p(t)$ and $m^p(t)$ during the interval $t\in[\tilde{t}^p_j,\tilde{t}^p_{j+1})$ along the solutions of \eqref{ctpe1}, \eqref{ctpe2}, \eqref{ggl}-\eqref{aaaae}, \eqref{obetbcp4f}-\eqref{tildeGqwrefa}. Suppose an event is triggered at $t=\tilde{t}^p_j$ and $m^p(\tilde{t}^p_j)>0$. Following the event at $t=\tilde{t}_j^p$, the control input is updated. Thus, we have from \eqref{mgw} that $\Gamma^p(\tilde{t}^p_j)=-\gamma m(\tilde{t}_j^p)-\frac{c}{\rho}W^p(\tilde{t}_j^p)<0$. Then, $\Gamma^p(t)$ will at least stay non-positive until $t=\tilde{t}^p_j+\tau$, where $\tau$ is the minimal dwell-time of the P-CETC (see R1 of Theorem \ref{exv}). This suggests that $\Gamma^p(t)$ remains non-positive in $t\in[\tilde{t}^p_j,\tilde{t}^p_j+h)$, since $h\leq\tau$. At every $t=nh,n>0,$ the periodic event-trigger \eqref{petwq}-\eqref{tildeGqwrefa} is evaluated, leading to an event trigger only if $\tilde{\Gamma}^p(nh)>0$, necessitating a control input update. In cases where $\tilde{\Gamma}^p(nh)\leq 0$, an update is not required as $\Gamma^p(t)$ remains non-positive due to \eqref{ineq}. This is because the RHS of \eqref{ineq} is definitely non-positive when $\tilde{\Gamma}^p(nh)\leq 0$. Thus, $\Gamma^p(t)$ will remain non-positive at least until $t=\tilde{t}_{j+1}^p$ where $\tilde{\Gamma}^p(\tilde{t}_{j+1}^{p-})> 0$. As $\Gamma^p(t)\leq 0$ for $t\in[\tilde{t}_{j}^p,\tilde{t}_{j+1}^p)$, we write from \eqref{petzq} and \eqref{mgw} that $d^2(t)\leq \gamma m^p(t)+\frac{c}{\rho}W^p(t)$ for $t\in[\tilde{t}_j^p,\tilde{t}_{j+1}^p)$. Then, considering the dynamics of $m^p(t)$ given by \eqref{ggl}, we get  $\dot{m}^p(t)\geq -(\eta+\gamma\rho)m^p(t)$ for $t\in(\tilde{t}^p_j,\tilde{t}^p_{j+1})$, which leads to $m^p(t)\geq e^{-(\eta+\gamma\rho)(t-\tilde{t}_j^p)}m^p(\tilde{t}_j^p)>0$ for $t\in[\tilde{t}_j^p,\tilde{t}_{j+1}^p)$. The time continuity of $m^p(t)$ leads to $m^p(\tilde{t}_{j+1}^{p-})=m^p(\tilde{t}_{j+1}^p)>0$. Therefore, after the control input has been updated at $t=\tilde{t}_{j+1}^p$, we obtain the equality $\Gamma^p(\tilde{t}_{j+1}^p)=-\gamma m^p(\tilde{t}^p_{j+1})-\frac{c}{\rho}W^p(\tilde{t}^p_{j+1})<0$. In a similar way, we can analyze the behavior of $\Gamma^p(t)$ and $m^p(t)$ in all $t\in[\tilde{t}^p_j,\tilde{t}^p_{j+1})$ for any $j\in\mathbb{N}$ starting from the first event at $\tilde{t}^p_0=0$ where  $m^p(0)>0$ to prove that $\Gamma^p(t)\leq 0$ for all $t\in[\tilde{t}^p_j,\tilde{t}^p_{j+1}),j\in\mathbb{N}$ and $m^p(t)>0$ for all $t>0$. As it holds that $m^p(t)>0$ for all $t>0$, the positive definiteness of $V^p(t)$ is confirmed. Thus, following similar arguments in the proof of Theorem \ref{exv}, it can be shown that $V^p(t)$ satisfies \eqref{zzzbnmk} for all $t\in(\tilde{t}_j^p,\tilde{t}_{j+1}^p),j\in\mathbb{N}$, and \eqref{asrt} for all $t>0$ as well as the global $L^2$-exponential convergence of the closed-loop system solution to zero satisfying the estimate \eqref{rrt},\eqref{MMm}. This completes the proof. \hfill $\square$

\subsubsection{P-STC}\label{sct_P_STC}
In this subsection, we develop the P-STC approach based on the P-CETC \eqref{obetbcp4a}-\eqref{aaaae}. This is accomplished by determining an upper bound on $d^2(t)$ and lower bound on $m^p(t)$, both of which are the constituent terms of the continuous-time event triggering function $\Gamma^p(t)$ given by \eqref{mgw}. We prove that updating the control input at specific times prescribed by the self-trigger ensures $\Gamma^p(t)$ remains non-positive and $m^p(t)$ remains positive along the P-STC closed-loop system solution. As a result, convergence properties equivalent to \textit{R4} and \textit{R5} in Theorem \ref{exv} holds for the P-STC closed-loop system as well.

Let $\check{I}^p =\{\check{t}^p_0,\check{t}^p_1,\check{t}^p_2,\ldots\}$ denote the sequence of event-times associated with the P-STC. Let the parameters $\gamma,c,\beta_1,\beta_2,\rho>0$ be selected as outlined in Section \ref{ass2}, and the parameter $\eta>0$ be selected as in \eqref{qptya}. The proposed P-STC strategy consists of two components:
\begin{enumerate}
    \item An event-triggered boundary control input 
    \begin{equation}\label{obetbpobwmngq}
\check{U}^p_{j}=\int_{0}^{1}k(y)u(y,\check{t}^p_{j})dy,
    \end{equation} for all $t\in[\check{t}_{j}^p,\check{t}^p_{j+1}),j\in\mathbb{N}$. Accordingly, the boundary condition \eqref{ctpe3} becomes
    \begin{equation}\label{zrt}
     u_{x}(1,t)+qu(1,t)=\check{U}^p_j.
    \end{equation}
       
\item A self-trigger determining the event-times
\begin{equation}\label{stss}
\check{t}_{j+1}^p=\check{t}^p_j+G^p(\check{t}_j^p),
\end{equation}
 with $\check{t}^p_0=0$ where $G^p(t)$ is a  uniformly and positively lower-bounded function defined as  \begin{equation}\label{cvbsq}
   \begin{split}
       &G^p(t) \triangleq\max\left\{\tau,\check{\tau}(t)\right\}.
    \end{split}
\end{equation}
Here, $\tau>0$ is given by \eqref{mdt} and
\begin{equation}\label{aacbnm}
\check{\tau}(t)\triangleq\frac{1}{2\lambda_{max}+\eta+c}\ln\left(\frac{\gamma m^p(t)+\frac{\gamma\rho H(t)}{2\lambda_{max}+\eta}+\frac{c}{\rho}W^p(t)}{H(t)+\frac{\gamma\rho H(t)}{2\lambda_{max}+\eta}}\right),
\end{equation}
where the dynamics of $m^p(t)$ satisfy \eqref{ggl}-\eqref{aaaae} along the solution of \eqref{ctpe1},\eqref{ctpe2},\eqref{obetbpobwmngq},\eqref{zrt} for $t\in(\check{t}_j^p,\check{t}_{j+1}^p),j\in\mathbb{N}$, and 
 \begin{equation}\label{aasder}
    H(t)  \triangleq 2\Vert k\Vert^2\Big(2+\frac{\varepsilon^2\Vert k\Vert^2}{\lambda^2_{max}}\Big)\Vert u[t]\Vert^2,
\end{equation}
with $k(y)$ given by \eqref{ase1}. 
\end{enumerate}

In the following lemma, we derive an upper bound on $d^2(t)$ and  a lower bound on $m^p(t)$ which are instrumental in proving the main results presented in Theorem \ref{hhgbsdz}.


\begin{lemm}\label{zbmmk} Consider the P-STC approach \eqref{obetbpobwmngq}-\eqref{aasder} under Assumption \ref{ass1}, which generates an increasing set of event times $\{\check{t}^p_{j}\}_{j\in\mathbb{N}}$ with $\check{t}^p_{j}=0$. Then, for the closed-loop system \eqref{ctpe1},\eqref{ctpe2},\eqref{obetbpobwmngq},\eqref{zrt} and the error $d(t)$ given by  \eqref{dt}, the following estimates hold
\begin{equation}\label{anx}
\begin{split}
    \Vert u[t]\Vert^2&\leq \Big(1+\frac{\varepsilon^2\Vert {k}\Vert^2}{\lambda^2_{max}}\Big)\Vert u[\check{t}_j^p]\Vert^2e^{2\lambda_{max}(t-\check{t}^p_j)},
\end{split}
\end{equation}
and
\begin{equation}\label{bbv111d}
    d^2(t)\leq H(\check{t}_j^p)e^{2\lambda_{max}(t-\check{t}_j^p)},
\end{equation}
for all $t\in[\check{t}^p_{j},\check{t}^p_{j+1}),j\in\mathbb{N}$ where $k(y)$ and $H(t)$ are given by \eqref{ase1} and \eqref{aasder}, respectively. Further, if the event-trigger parameters $\gamma,c,\beta_1,\beta_2,\rho>0$ are chosen as outlined in Section \ref{ass2}, and $\eta>0$ is chosen as in \eqref{qptya}, then $W^p(t)$ given by \eqref{aacbnmh}
satisfies the following relations
\begin{equation}\label{akpgh}
    W^p(t)\geq e^{-(b^*+c)(t-\check{t}_j^p)}W^p(\check{t}_j^p)\text{ with }b^*=\eta,
\end{equation}
 for all $t\in\big[\check{t}_j^p,\check{t}_{j+1}^p\big),j\in\mathbb{N},$ and
\begin{equation}\label{akpqgh}
   W^p(t)\geq 0,\text{ i.e., } V^p(t)\leq e^{-b^*t}V_0\text{ with }b^*=\eta,
\end{equation}
for all $t>0$ whereas $m^p(t)$ governed by \eqref{ggl}-\eqref{aaaae} satisfies
\begin{equation}\label{bbv211}
\begin{split}
    m^p(t)\geq& m^p(\check{t}_j^p)e^{-\eta(t-\check{t}_j^p)}\\&-\frac{\rho H(\check{t}_j^p)}{2\lambda_{max}+\eta}e^{-\eta (t-\check{t}_j^p)}\Big(e^{(2\lambda_{max}+\eta)(t-\check{t}_j^p)}-1\Big),
\end{split}
\end{equation}
 for all $t\in[\check{t}^p_{j},\check{t}^p_{j+1}),j\in\mathbb{N}$.
\end{lemm}

\noindent\textbf{Proof} Consider the positive definite function
\begin{equation}\label{fffg}
    E(t) = \frac{1}{2}\int_{0}^{1}u^2(x,t)dx.
\end{equation}
Taking its time derivative  along the solution of \eqref{ctpe1},\eqref{ctpe2},\eqref{obetbpobwmngq},\eqref{zrt} and using Young's inequality, we show that
\begin{equation}\label{ddsa}
    \begin{split}
        &\dot{E}(t)\leq -\varepsilon qu^2(1,t)+\lambda_{max}\Vert u[t]\Vert^2+\frac{\varepsilon \nu}{2} u^2(1,t)+\frac{\varepsilon}{2\nu}(\check{U}_j^p)^2,
    \end{split}
\end{equation}
for all $t\in(\check{t}^p_{j},\check{t}^p_{j+1}),j\in\mathbb{N}$ and some $\nu>0$. Let us select $\nu$ as $\nu=\lambda_{max}/2\varepsilon$. Then, one can rewrite \eqref{ddsa} as
\begin{equation}\label{sssf}
\begin{split}
    \dot{E}(t)\leq& -\varepsilon\Big( q-\frac{\lambda_{max}}{4\varepsilon}\Big)u^2(1,t)+\lambda_{max}\Vert u[t]\Vert^2+\frac{\varepsilon^2}{\lambda_{max}}(\check{U}^p_j)^2,
\end{split}
\end{equation}
for $t\in(\check{t}^p_{j},\check{t}^p_{j+1}),j\in\mathbb{N}$. Using Cauchy-Schwarz inequality on \eqref{obetbpobwmngq} and considering \eqref{fffg}, we obtain $(\check{U}^p_j)^2\leq 2\Vert k\Vert^2E(\check{t}^p_j).$ Thus, recalling Assumption \ref{ass1} from which it follows that $q>\lambda_{max}/4\varepsilon$, we write \eqref{sssf} as $\dot{E}(t)\leq 2\lambda_{max} E(t)+\frac{2\varepsilon^2\Vert k\Vert^2}{\lambda_{max}}E(\check{t}_j^p),$
for $t\in(\check{t}_{j}^p,\check{t}^p_{j+1}),j\in\mathbb{N}$. As a result, we show that $E(t)\leq e^{2\lambda_{max}(t-\check{t}_j)}E(\check{t}^p_j)+\frac{\varepsilon^2\Vert k\Vert^2}{\lambda^2_{max}}E(\check{t}_j^p)\Big(e^{2\lambda_{max}(t-\check{t}_j^p)}-1\Big),$ for $t\in[\check{t}_j^p,\check{t}_{j+1}^p),j\in\mathbb{N}$ from which we obtain \eqref{anx}. By applying the Cauchy-Schwarz inequality and Young's inequality to \eqref{dt} over the interval $t\in[\check{t}_j^p,\check{t}_{j+1}^p)$,  we get $ d^2(t)\leq 2\Vert k\Vert^2\Vert u[\check{t}^p_j]\Vert^2+2\Vert k\Vert^2\Vert u[t]\Vert^2.$ Then, using \eqref{anx}, we obtain \eqref{bbv111d}. As the event-trigger parameters $\gamma,\beta_1,\beta_2,\rho$ are selected as outlined in Section \ref{ass2}, and $\eta>0$ is selected as in \eqref{qptya}, we obtain the relations \eqref{akpgh} and \eqref{akpqgh} following the same arguments in Lemma \ref{aqrt}. Thus, Considering the dynamics of $m^p(t)$ given by \eqref{ggl}-\eqref{aaaae} and the relations \eqref{bbv111d} and \eqref{akpqgh}, we show that $\dot{m}^p(t)\geq -\eta m^p(t)-\rho H(\check{t}_j^p)e^{2\lambda_{max}(t-\check{t}_j^p)},$
for $t\in(\check{t}_{j}^p,\check{t}_{j+1}^p),j\in\mathbb{N}$ from which we obtain \eqref{bbv211}.  This completes the proof. \hfill $\square$


\begin{thme}\textnormal{\textbf{(Results under P-STC)}}\label{hhgbsdz} Consider the P-STC approach \eqref{obetbpobwmngq}-\eqref{aasder} under Assumption \ref{ass1}, which generates an increasing set of event-times $\check{I}^p=\{\check{t}_j^p\}_{j\in\mathbb{N}}$ with $\check{t}_0^p=0$. For every $u[0]\in L^{2}(0,1)$, there exists a unique solution $u:\mathbb{R}_+\times[0,1]\rightarrow\mathbb{R}$ such that $u\in C^0(\mathbb{R}_+;L^2(0,1)\cap C^1(\check{J}^p\times [0,1])$ with $u[t]\in C^2([0,1])$ which satisfy \eqref{ctpe2},\eqref{obetbpobwmngq},\eqref{zrt} for all $t>0$ and \eqref{ctpe1} for all $t>0,x\in(0,1),$ where $\check{J}^p=\mathbb{R_{+}}\text{\textbackslash}\check{I}^p$. If the event-trigger parameters $\gamma,c,\beta_1,\beta_2,\rho>0$ are chosen as outlined in Section \ref{ass2}, and $\eta>0$ is selected as in \eqref{qptya}, then the following results hold:
\begin{enumerate}
    \item[R1:] $\Gamma^p(t)$ given by \eqref{mgw}  satisfies $\Gamma^p(t)\leq 0$ along the solution of \eqref{ctpe1},\eqref{ctpe2},\eqref{ggl}-\eqref{aaaae},\eqref{obetbpobwmngq}-\eqref{aasder} for all $t>0$.  
    \item[R2:] The dynamic variable $m^p(t)$ governed by \eqref{ggl}-\eqref{aaaae} with $m^p(0)=m^r(0)>0$ satisfies $m^p(t)>0$ along the solution of \eqref{ctpe1},\eqref{ctpe2},\eqref{obetbpobwmngq}-\eqref{aasder} for all $t>0$.
\item[R3:] The Lyapunov candidate $V^p(t)$ given by \eqref{abx},\eqref{asf} satisfies \eqref{zzzbnmk} for all $t\in (\check{t}_j^p,\check{t}_{j+1}^{p}),j\in\mathbb{N}$ and \eqref{asrt} for all $t>0$, with $b^*=\eta$. 
    
    \item[R4:] The closed-loop solution of \eqref{ctpe1},\eqref{ctpe2},\eqref{obetbpobwmngq}-\eqref{aasder} globally exponentially converges to zero in the spatial   $L^2$ norm satisfying the estimate \eqref{rrt},\eqref{MMm} with $b^*=\eta$.
\end{enumerate}
\end{thme}

\begin{figure*}
\centering
\includegraphics[scale=0.675]{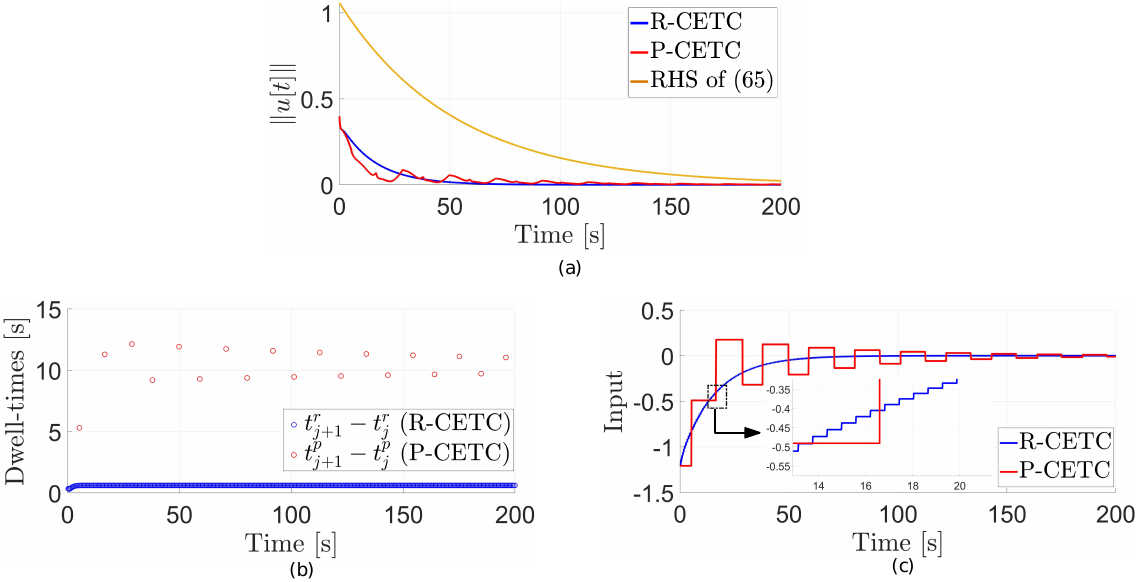}
\caption{Results under the R/P-CETC.}
\label{acnw1dgh}
\end{figure*}

\textit{Proof.} The well-posedness of the closed-loop system \eqref{ctpe1}, \eqref{ctpe2}, \eqref{obetbpobwmngq},\eqref{zrt} is a direct consequence of Proposition \ref{cor1}. The system's solution for all $t>0$ is obtained by iteratively applying Proposition \ref{cor1} between events. Given that the parameters $\gamma, c, \beta_1, \beta_2, \rho>0$ are chosen in accordance with Section \ref{ass2} and $\eta>0$ is set as per \eqref{qptya}, Lemma \ref{zbmmk} ensures that $W^p(t)\geq 0$ for all $t>0$ under the P-STC framework \eqref{obetbpobwmngq}-\eqref{aasder}. Let us consider the function $\Gamma^p(t)$ given by \eqref{mgw}, along the trajectories \eqref{ctpe1}, \eqref{ctpe2}, \eqref{ggl}-\eqref{aaaae}, \eqref{obetbpobwmngq}-\eqref{aasder}. If an event occurs at time $t=\check{t}_j^p$ and $m^p(\check{t}_j^p)>0$, the subsequent control input update ensures, as per \eqref{mgw}, that $\Gamma^p(\check{t}_j^p)=-\gamma m^p(\check{t}_j^p)-\frac{c}{\rho}W^p(\check{t}_j^p)<0$. Moreover, $\Gamma^p(t)$ stays non-positive at least until the time $t=\check{t}_j^p+\tau$, with $\tau$ being the minimal dwell-time of the P-CETC (refer to R1 of Theorem \ref{exv}). We have from \eqref{bbv111d} that $d^2(t)\leq H(\check{t}_j^p)e^{2\lambda_{max}(t-\check{t}_j^p)}$ and from \eqref{akpgh} and \eqref{bbv211} that
\begin{equation*}\label{bbv211q}
\begin{split}
    &\gamma m^p(t)+\frac{c}{\rho} W^p(t) \geq \gamma m^p(\check{t}_j^p)e^{-\eta(t-\check{t}_j^p)}+\frac{c}{\rho}e^{-(\eta+c)(t-\check{t}_j^p)}W^p(\check{t}_j^p)\\&\qquad\qquad -\frac{\gamma\rho H(\check{t}_j^p)}{2\lambda_{max}+\eta}e^{-\eta (t-\check{t}_j^p)}\Big(e^{(2\lambda_{max}+\eta)(t-\check{t}_j^p)}-1\Big),\\&\geq \gamma m^p(\check{t}_j^p)e^{-(\eta+c)(t-\check{t}_j^p)}+\frac{c}{\rho}e^{-(\eta+c)(t-\check{t}_j^p)}W^p(\check{t}_j^p)\\&\quad -\frac{\gamma\rho H(\check{t}_j^p)}{2\lambda_{max}+\eta}e^{2\lambda_{max}(t-\check{t}_j^p)}+\frac{\gamma\rho H(\check{t}_j^p)}{2\lambda_{max}+\eta}e^{-(\eta+c)(t-\check{t}_j^p)},
\end{split}
\end{equation*}
for $t\in[\check{t}_j^p,\check{t}_{j+1}^p)$. Note that the RHS of \eqref{bbv111d} is an increasing function of $t$ whereas the RHS of \eqref{bbv211} is a decreasing function of $t$. Then, if there exists a positive solution $t^\dagger>\check{t}_j^p$ that satisfies   
\begin{equation}\label{alfg}
\begin{split}
&H(\check{t}_j^p)e^{2\lambda_{max}(t^\dagger-\check{t}_j^p)}\\&=
\gamma m^p(\check{t}_j^p)e^{-(\eta+c)(t^\dagger-\check{t}_j^p)}+\frac{c}{\rho}e^{-(\eta+c)(t^\dagger-\check{t}_j^p)}W^p(\check{t}_j^p)\\&\quad -\frac{\gamma\rho H(\check{t}_j^p)}{2\lambda_{max}+\eta}e^{2\lambda_{max}(t^\dagger-\check{t}_j^p)}+\frac{\gamma\rho H(\check{t}_j^p)}{2\lambda_{max}+\eta}e^{-(\eta+c)(t^\dagger-\check{t}_j^p)},
\end{split}
\end{equation}
we are certain that $d^2(t)\leq \gamma m^p(t)+\frac{c}{\rho}\big(e^{-b^*t}V_0^p-V^p(t)\big)$ \textit{i.e.,} $\Gamma^p(t)\leq 0$ for $t\in[\check{t}_j^p,t^\dagger)$. This is because the LHS of \eqref{alfg} is an upper bound on $d^2(t^\dagger)$, and the RHS  of \eqref{alfg} is a lower bound on $\gamma m^p(t^\dagger)+\frac{c}{\rho}W^p(t^\dagger)$. Solving \eqref{alfg} for $t^\dagger$, we obtain that $t^\dagger = \check{t}_j^p+\check{\tau}(\check{t}_j^p)$, where $\check{\tau}(t)$ is given by \eqref{aacbnm}. If \( t^\dagger > \check{t}_j^p + \tau \), the next event is set as \( \check{t}_{j+1}^p = t^\dagger \). If \( t^\dagger \leq \check{t}_j^p + \tau \), then the next event is set as \( \check{t}_{j+1}^p = \check{t}_j^p + \tau \). Since the next event time is given by \eqref{stss},\eqref{cvbsq}, it is ensured that \( \Gamma^p(t) \leq 0 \) for \( t \in [\check{t}^p_j, \check{t}^p_{j+1}) \) while preventing the Zeno phenomenon. Using the same reasoning as in the proof of Theorem \ref{cvgh}, we show that \( m^p(t) > 0 \) for all \( t > 0 \), confirming the positive definiteness of \( V^p(t) \). Therefore, similar arguments in the proof of Theorem \ref{exv} show that \( V^p(t) \) satisfies \eqref{zzzbnmk} for all \( t \in (\check{t}_j^p, \check{t}_{j+1}^p) \), \( j \in \mathbb{N} \), and \eqref{asrt} for all \( t > 0 \), ensuring the global \( L^2 \)-exponential convergence of the closed-loop system solution to zero, satisfying the estimate \eqref{rrt}, \eqref{MMm}. This completes the proof.
\hfill $\square$


\begin{rmk}\rm\label{aannmmpp}
    In \cite{rathnayake2023observertac}, the authors develop observer-based PETC and STC strategies equipped with strictly decreasing Lyapunov functions, for a class of reaction-diffusion PDEs. We refer to them as regular PETC (R-PETC) and regular STC (R-STC) to differentiate them from the P-PETC and P-STC introduced in Sections \ref{sct_P_PETC} and \ref{sct_P_STC}, respectively. It is important to underscore that the  P-PETC and P-STC proposed here generalize the full-state feedback versions of the R-PETC and R-STC proposed in \cite{rathnayake2023observertac} by allowing for Lyapunov functions that are not decreasing at all times. By setting $c=0$, one can recover the R-PETC and R-STC from the corresponding P-PETC and P-STC.  
\end{rmk}

\begin{rmk}\rm
For the R/P-CETC, R-PETC, and R-STC, the event-trigger parameter $\eta>0$ can be freely selected (see Theorem \ref{exv} and the reference \cite{rathnayake2023observertac}). However, for both P-PETC and P-STC, $\eta>0$ should be chosen such that $\eta\leq 2b/B$, as specified by \eqref{qptya}. Some readers might question if this upper bound constraint on $\eta$ places the P-PETC and P-STC at a disadvantage compared to the R-PETC and R-STC. However, we would like to emphasize that increasing $\eta$ beyond $2b/B$ does not enhance the theoretical convergence guarantee (performance-barrier). This is because the exponential decay $b^*$ in the estimate \eqref{hhf} and \eqref{asrt} is given by $b^*=\min\big\{\frac{2b}{B},\eta\big\}$ (see \eqref{BBB}). Thus, for $\eta>\frac{2b}{B}$, the decay rate remains $b^*=\frac{2b}{B}$, regardless of how large $\eta$ becomes.
\end{rmk}

\section{Numerical Simulations}

\begin{figure*}
\centering
\includegraphics[scale=0.675]{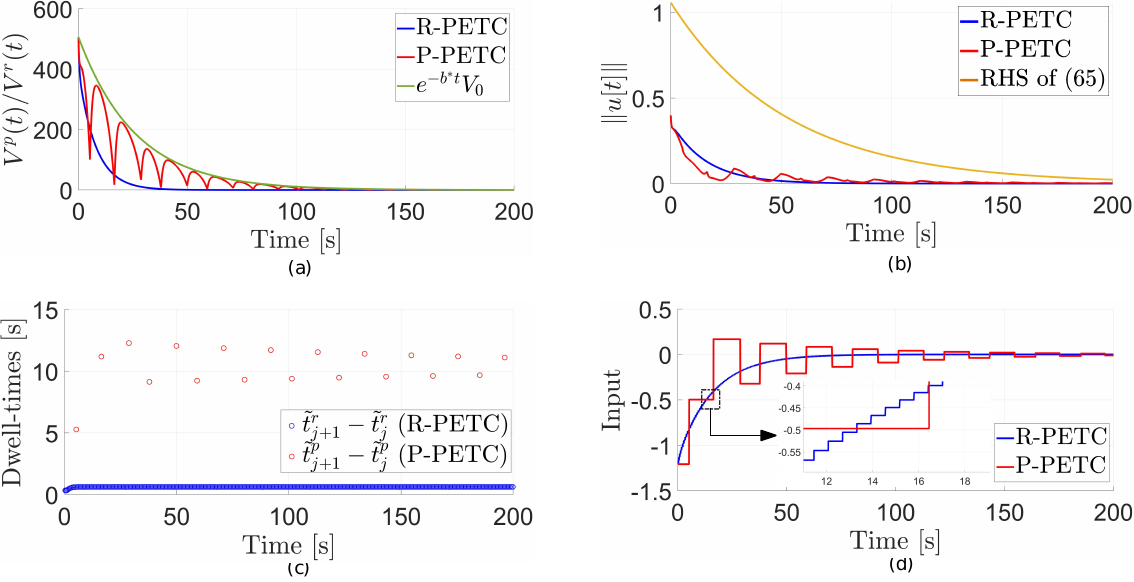}
\caption{Results under the R/P-PETC.}
\label{acnw1qac}
\end{figure*}

\begin{figure*}
\centering
\includegraphics[scale=0.675]{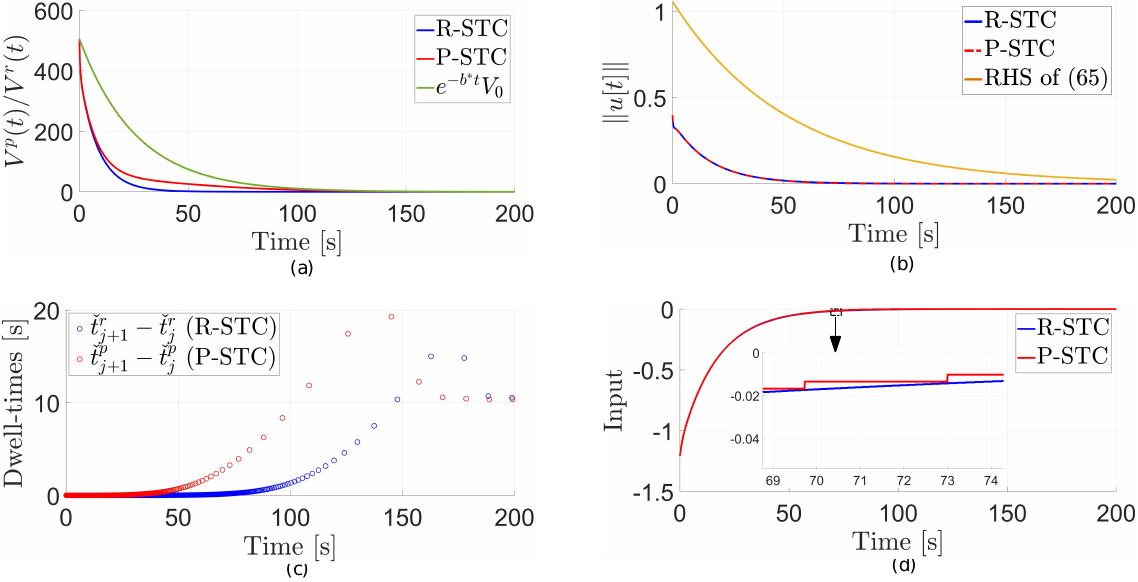}
\caption{Results under the R/P-STC.}
\label{acnw1mnbv}
\end{figure*}

To illustrate the efficacy of the proposed designs, we consider a reaction-diffusion system with constant parameters $\varepsilon=0.1,\lambda(x)\equiv \lambda_{max}=\lambda=0.25,q=2,\theta_1=1,\theta_2=0,$ and the initial conditions $u[0]=10x^2(x-1)^{2}$. The gain kernels \eqref{ctcke1}-\eqref{aamlper} and \eqref{ctckeg}-\eqref{kkllspip} have explicit solutions, and readers are referred to \cite{rathnayake2021observer} for details. The parameters for the event-triggers are chosen as follows: $m(0)=10^{-4}$, and  $\sigma=0.9$. It can be shown using \eqref{al1},\eqref{al2} that $\alpha_{1}= 0.3466, \alpha_2= 0.5405$. Therefore, from \eqref{betas}, we obtain $\beta_{1}=3.4665, \beta_{2}= 5.4055$. Let us choose $B$ and $\kappa$ as $B= 3308.7$ and $\kappa=5$ so that \eqref{Bs} is satisfied. Then, from \eqref{hhjknbv}, we obtain $\rho=827.1872$. We choose $\eta=0.0383$ so that \eqref{qptya} is satisfied. The computed minimal dwell-time is $0.01$s. Thus, we use $\Delta t=0.001$s to time discretize the plant dynamics using the implicit Euler scheme. We set $h=0.01$s as the sampling period for the R/P-PETC approaches. Space discretization is done using a step size of $\Delta x=0.005$.  We use $c=1$ for the P-CETC and P-PETC for generating Figs. \ref{acnw1dgh} and \ref{acnw1qac}, whereas we use $c=0.01$ for the P-STC for generating Fig. \ref{acnw1mnbv}. The R-CETC, R-PETC, and R-STC results are generated by setting $c=0$ in the corresponding P-ETC design, as mentioned in Remark \ref{aannmmpp}.  

In Figs. \ref{acnw11}-\ref{acnw1mnbv}, we compare the R-ETC to the P-ETC under continuous-time event-triggered, periodic event-triggered, and self-triggered configurations. From Fig. \ref{acnw11}, Fig. \ref{acnw1qac}(a), and Fig. \ref{acnw1mnbv}(a), we can observe that the Lyapunov functions under the P-CETC, P-PETC, and P-STC are above their regular counterparts R-CETC, R-PETC, and R-STC, respectively. This is while still respecting the performance-barrier $e^{-b^*t}V_0$, indicating the flexibility of the performance-barrier-based approach. Notably, Lyapunov functions under the P-CETC and P-PETC sometimes even increase while respecting the performance-barrier. Consequently, we can observe in Fig. \ref{acnw1dgh}(b) and Fig. \ref{acnw1qac}(c) that the dwell-times under the P-CETC and P-PETC are significantly larger than those under the R-CETC and R-PETC, respectively. Similarly, Fig. \ref{acnw1dgh}(c) and Fig. \ref{acnw1qac}(d) show that the P-CETC and P-PETC result in less frequent control updates than the R-CETC and R-PETC. In Fig. \ref{acnw1mnbv}(c), the dwell-times under the P-STC start larger than the R-STC, but as time progresses, the dwell-times for both approaches converge to similar values. From Fig. \ref{acnw1dgh}(a) and Fig. \ref{acnw1qac}(b), we can observe that the closed-loop signal $\Vert u[t]\Vert$ under the P-CETC and P-PETC takes longer to converge than the R-CETC and R-PETC, respectively, whereas $\Vert u[t]\Vert$ under the P-STC and R-STC converges to zero at similar rates. Nevertheless, it is crucial to emphasize that performance-barrier-based designs converge faster than the performance-barrier itself. The closed-loop signal $\Vert u[t]\Vert$ under these designs adheres to the decay estimate given by equations \eqref{rrt},\eqref{MMm}. \begin{table*}
\centering
\begin{tabular}{|l|r|r|r|r|r|r|r|}
\hline
\multicolumn{1}{|c}{} & \multicolumn{1}{|c|}{Regular}           & \multicolumn{6}{c|}{Performance-Barrier} \\
\hline
     & c=0    & c=0.001 & c=0.01  & c=0.1   & c=1     & c=10    & c=100   \\  \hline
CETC Avg. DTs [s] & 0.6104 & 6.8585  & 9.596   & 11.9984 & 10.3245 & 10.3261 & 10.3422 \\ \hline
PETC Avg. DTs [s] & 0.6178 & 7.077   & 9.7182  & 12.391  & 10.3325 & 10.3258 & 10.3387 \\ \hline
STC Avg. DTs [s]  & 0.1037 & 0.1433  & 0.2369  & 0.4269  & 0.3682  & 0.1366  & 0.05    \\ \hline
\end{tabular}
\caption{Average dwell-times (Avg. DTs) in seconds over a period of $500s$. The case $c=0$ corresponds to the R-ETC, whereas the cases $c\in\{0.001,0.01,0.1,1,10,100\}$ corresponds to the P-ETC. Note that the average dwell-time does not fully depict the nature of the dwell-times. It is only intended to show of how increasing $c$ affects dwell-times.}
\label{hjklpi}
\end{table*}
In Table \ref{hjklpi}, we show the average dwell-times for the proposed strategies over a period of $500s$. It is worth noting that the average dwell-time does not fully depict the nature of the dwell-times. It is intended to give the reader an idea of how increasing $c$ affects dwell-times. We observe that under the P-CETC and P-PETC, the average dwell-time over $500s$ increases as $c$ increases and tends to saturate around $10s$. Under the P-STC, the average dwell-time over $500s$ initially increases with $c$ and then starts to decrease as $c$ increases further. This can be understood via \eqref{aacbnm}, where the factor $1/(2\lambda+\eta+c)$ causes the dwell-time to decrease as $c$ increases beyond a certain threshold.

\section{Conclusions}
This paper has developed a new approach for event-triggered boundary control, called performance-barrier event-triggered control, for a class of reaction-diffusion PDEs. The core of this strategy lies in allowing for the Lyapunov function of the closed-loop system to diverge from a strictly monotonic decrease as long as it remains below an acceptable performance-barrier.  While ensuring that this performance is met, this novel method can result in longer dwell times between events compared to regular methods that enforce a constant decrease in the Lyapunov function. 

The concept of performance residual--the difference between the value of the performance-barrier and the Lyapunov function--plays a crucial role in achieving these results. Integrating this concept into the triggering mechanism allows for enhanced flexibility in the behavior of the Lyapunov function. It is also proven that the new performance-barrier event-triggered control strategy ensures global exponential convergence of the closed-loop system solution to zero in the spatial $L^2$ norm and guarantees a Zeno-free behavior. Furthermore, we have successfully extended the performance-barrier event-triggered control method to its periodic event-triggered and self-triggered variants. These variants, designed to avoid the need for continuous monitoring of the event-trigger, are shown to provide performance equivalent to that of their continuous-time event-triggered counterpart. We have conducted numerical simulations that illustrate the performance of the proposed control designs. A natural direction for future research is the design of observer-based variants of performance-barrier-based designs.

 \bibliographystyle{IEEEtranS}
\bibliography{main}

\end{document}